%% file: main.tex
\documentclass[lettersize,journal]{IEEEtran}
\input{header}

\begin{document}


\title{\sysname: LLM-based Anomaly Detection for \\ Spatial-Temporal Graph in Industrial Systems}

\author{Yuchen Zhang, 
Ning Xi,~\IEEEmembership{Member,~IEEE,}, Pengbin Feng, Shigang Liu,~\IEEEmembership{Member,~IEEE,} \\ Jianfeng Ma,~\IEEEmembership{Member,~IEEE,} Yulong Shen,~\IEEEmembership{Member,~IEEE,} Yanan Sun, and Xiaolin Zhou
\thanks{Y. Zhang, N. Xi, P. Feng, J. Ma, Y. Sun, and X. Zhou are with the School of Cyber Engineering, Xidian University, Xi'an 710126, China. E-mail: zhangyuchen@stu.xidian.edu.cn, nxi@xidian.edu.cn, pbfeng@xidian.edu.cn, jfma@mail.xidian.edu.cn, syn2001919@163.com, zhouxiaolin0223@163.com}
\thanks{S. Liu is with the School of Science, Computing and Engineering Technologies, Swinburne University of Technology, Hawthorn, VIC, Australia. E-mail: shigangliu@swin.edu.au}
\thanks{Y. Shen is with the School of Computer Science and Technology, Xidian University, Xi'an 710126, China. E-mail: ylshen@mail.xidian.edu.cn}
}



\maketitle

\begin{abstract}
\input{tex/Abstract}
\end{abstract}


\section{INTRODUCTION}
\input{tex/Introduction}

\section{PRELIMINARIES AND THREAT MODEL}
\input{tex/Preliminaries}

\section{PROPOSED \sysname}
\input{tex/Approach}

\section{EXPERIMENTS AND EVALUATION}
\input{tex/ExperimentsEvaluation}

\section{DISCUSSION}
\input{tex/Discussion}

\section{RELATED WORK}
\input{tex/RelatedWork}

\section{CONCLUSION}
\input{tex/Conclusion}

\bibliographystyle{IEEEtran}
\bibliography{sample.bib}

\end{document}

%% file: header.tex
\usepackage{amsmath,amsfonts}
\usepackage{array}
\usepackage[caption=false]{subfig}
\usepackage{textcomp}
\usepackage{stfloats}
\usepackage{url}
\usepackage{verbatim}
\usepackage{cite}
\usepackage{caption}
\usepackage{xspace}
\usepackage{svg}
\usepackage{float} 
\usepackage{graphicx}
\usepackage{filecontents}
\usepackage[ruled,vlined,linesnumbered]{algorithm2e}
\usepackage{cite}
\usepackage{url}
\usepackage{xspace}
\usepackage{amssymb}
\usepackage{booktabs}
\usepackage{multirow}
\usepackage{makecell}
\usepackage{diagbox}
\usepackage{tabularx}
\usepackage[table]{xcolor}

\newcommand{\sysname}{\textsc{IstGPT}\xspace}
\newcommand{\llm}{\textsc{IIOT-LLM}\xspace}
\newcommand{\tool}[1]{{\textit{#1}}}

%% file: tex/Abstract.tex
Industrial Internet systems face increasing threats from sophisticated industrial control system (ICS) attacks, resulting in critical safety incidents. However, existing tools exhibit limited effectiveness in real-time anomaly detection due to the complex dependencies among sensors and actuators. To tackle this, we present \sysname, the first industrial anomaly detection tool based on LLMs and graph learning to provide real-time protection against a wide range of ICS attacks.
\sysname achieves fine-grained and precise modeling on spatial-temporal dependencies in industrial cyber-physical systems.
It first leverages industrial multi-modal knowledge, including operational data, technical documents, and system diagrams, to extract sensor-actuator dependency graphs via multi-stage prompt engineering. Then, LLM-Optimation iteratively refines the graph based on node accuracy, edge consistency, and logical coherence. Finally, \sysname integrated improved graph neural networks with an encoder-decoder architecture to detect anomalies via reconstruction errors.
We evaluate \sysname against 12 state-of-the-art baselines on 9 datasets, including 2 public, 6 simulated, and a real-world robotic arm dataset. \sysname achieves the best F1-scores and eTaF1 (a newer time-aware metric) across nine datasets.
We further discuss the feasibility of deploying \sysname in real-world industrial scenarios.

%% file: tex/Introduction.tex

\IEEEPARstart{I}{ndustrial} Internet has emerged as a driving force for the evolution of next-generation industrial systems and the advancement of human society. It employs sensing, communication, computing, and control technologies for automatic and intelligent manufacturing, energy management, etc.~\cite{hu2024industrial}

However, Industrial Control Systems (ICS) have suffered multiple high-impact cyber attacks \cite{buchanan2022cyber, salazar2024tale, pollard2024case, cervini2022don, liu2025intrusion}, exposing their security fragility.
Attackers compromise cyber and physical layers of industrial systems, leading to unauthorized manipulation of sensors and actuators. These ICS attacks disrupt the normal operation of industrial equipment, potentially resulting in critical safety incidents.



Previous detection techniques, including network intrusion detection \cite{hu2021deep}, control behavior detection \cite{ike2023scaphy}, and anomaly detection \cite{nassif2021machine}, are used to detect ICS attacks. 
Among these techniques, anomaly detection stands out for its ability to detect complex zero-day ICS attacks and adaptability to various industrial scenarios. As a result, industrial anomaly detection has garnered significant attention from academia and industry. 

\begin{figure}[tbp]
\centering
\includegraphics[width=0.95\linewidth]{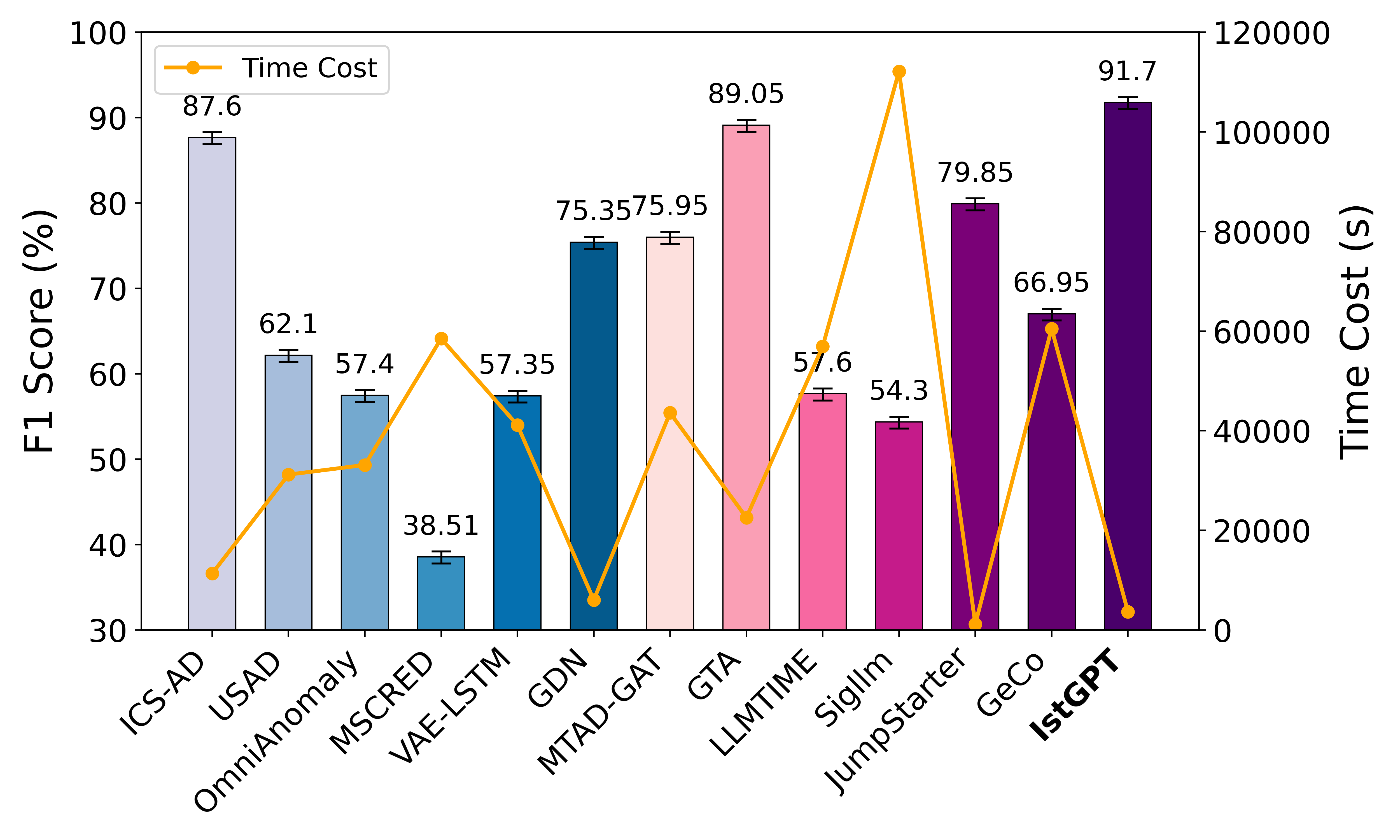}
\caption{\label{fig_comp_pub} Overall Performance Evaluation of Existing Industrial Anomaly Detection Methods on Public Datasets (SWaT\&WADI).}
\end{figure}

Based on our empirical analysis of prior work~\cite{lamberts2023sok, giraldo2018survey, luo2021deep, apruzzese2023sok, balla2022applications, fung2022perspectives, qin2023mtad, fung2024attributions, audibert2020usad, su2019robust, zhang2019deep, lin2020anomaly, deng2021graph, zhao2020multivariate, chen2021learning, gruver2023large, alnegheimish2024large, ma2021jump, zhan2022stgat}, most industrial anomaly detection methods fall into two categories: (1) \tool{traditional rule-based methods}, which detect anomalies by comparing observations against fixed reconstruction patterns or predefined control logic, and (2) \tool{learning-based methods}, further divided into \tool{sequence representation learning-based} and \tool{graph learning-based} approaches. The former learns temporal representations of sensor-actuator sequences to model normal behavior, while the latter employs graph neural networks to capture data-driven dependencies for anomaly detection. A major reason for the limited performance of existing anomaly detection methods is the lack of effective modeling of dependencies among industrial sensors and actuators. Due to cyber-physical connections, variables in real-world industrial systems exhibit well-structured underlying topologies \cite{wu2023physics}. \tool{Traditional rule-based methods} rely on handcrafted rules or static configurations \cite{chi2022knowledge,ike2023scaphy, lin2018tabor}, which are often limited in scalability and generalizability. \tool{Learning-based methods} construct dependency graphs purely from process data, tend to be inconsistent and often include spurious or misleading relationships \cite{shuaiyi2023process, wu2023physics}.

The performance of existing methods across different industrial scenarios and diverse ICS attack types remains limited. We evaluated 12 state-of-the-art (SOTA) baselines on two public datasets, using F1-score and time cost as metrics, as shown in Fig.~\ref{fig_comp_pub}. Most methods achieve detection performance below 80\%. Moreover, the time cost of training remains high even on datasets containing as few as fifty to hundreds of nodes.

In practice, industrial prior knowledge contains rich physical semantics extending beyond process data alone \cite{karniadakis2021physics,wu2022data}. Recently, large language models (LLMs) have demonstrated remarkable capabilities in multi-modal knowledge processing \cite{hurst2024gpt}. LLMs can extract temporal and procedural dependencies among sensors and actuators from diverse industrial knowledge, including system design documents, process data, and testbed diagrams. 
Therefore, novel methodologies are needed to leverage LLMs for constructing accurate, interpretable sensor-actuator dependencies from industrial multi-modal knowledge. To address this, two key challenges must be overcome.

\textbf{Challenge \#1: Lack of Iterative Prompt Engineering for Constructing Complex Sensor-Actuator Dependency Graphs in Industrial Systems.} 
Most existing LLM-based methods in industrial scenarios frame time-series forecasting as a next-token prediction task by encoding numerical values as text \cite{gruver2023large,alnegheimish2024large}. However, these methods primarily focus on univariate time series and overlook the relationships among sensors and actuators.
There is a lack of research on prompt engineering strategies tailored to guide LLMs in understanding industrial multi-modal knowledge and translating it into meaningful dependency structures.
Moreover, no classification schema systematically defines industrial multi-modal knowledge to support prompt-driven dependency graph construction.
Therefore, the first challenge is to design an effective prompt engineering that guides LLMs to construct sensor-actuator dependency graphs from classified industrial multi-modal knowledge.

\textbf{Challenge \#2: Critical Need for Evaluation Rules and Metrics in LLM-Based Dependency Graph Construction.} 
Purely data-driven methods often yield physically inconsistent graphs with spurious relationships, degrading anomaly detection performance \cite{wu2023physics}.
A key reason for the inaccuracy of such dependency graph construction is the absence of rule-based evaluation metrics to validate the quality of the generated graphs. 
Without these validation rules and feedback mechanisms, the generated graphs are often incomplete or incorrect, failing to capture the true cyber-physical interactions within industrial systems.
Therefore, the second challenge is to develop evaluation criteria supporting the assessment and iterative refinement of LLM-generated sensor-actuator dependency graphs.

\textbf{Our solution:} To overcome the above challenges, we propose a novel anomaly detection approach for industrial cyber-physical systems based on LLMs, called \sysname. \sysname leverages LLMs to construct a sensor-actuator spatial dependency graph and employs deep learning methods to model spatial-temporal patterns, including four phases.
(1) \sysname preprocesses raw industrial sensor and actuator data through cleaning, normalization, and sliding-window segmentation to ensure structured and reliable inputs.
(2) \sysname constructs an LLM-based sensor-actuator dependency graph using multi-stage prompt engineering over multi-modal industrial knowledge (e.g., operational data, technical documents, and diagrams). The graph integrates temporal correlations to form an Industrial Spatial-Temporal Graph (ISTG). Graph construction completes within 7 minutes on average for industrial scenarios with hundreds of nodes, demonstrating practical feasibility.
(3) \sysname applies unsupervised machine learning, including GAT, GCN, and an encoder–decoder architecture, to model normal spatio-temporal behaviors. 
(4) \sysname detects and localizes anomalies by thresholding reconstruction errors, achieving efficient detection with low time cost.
We evaluate \sysname on nine datasets, including two public benchmarks and seven private industrial datasets. Compared with 12 state-of-the-art baselines (e.g., \tool{ICS-AD}~\cite{fung2024attributions}, \tool{USAD}~\cite{audibert2020usad}, \tool{OmniAnomaly}~\cite{su2019robust}, \tool{MSCRED}~\cite{zhang2019deep}, \tool{VAE-LSTM}~\cite{lin2020anomaly}, \tool{GDN}~\cite{deng2021graph}, \tool{MTAD-GAT}~\cite{zhao2020multivariate}, \tool{GTA}~\cite{chen2021learning}, \tool{LLMTIME}~\cite{gruver2023large}, \tool{Sigllm}~\cite{alnegheimish2024large}, \tool{JumpStarter}~\cite{ma2021jump}, and \tool{GeCos}~\cite{wolsing2025gecos}), \sysname achieves an average F1-score of 91.7\% on public datasets, outperforming the top baselines (89.1\%, 87.6\%, and 79.9\%). It also demonstrates superior performance on all private datasets from industrial simulations and a real-world robotic arm.
Our contributions are summarized as follows:
\begin{itemize}
\item We are the first to design an iterative prompt engineering to capture complex sensor-actuator dependencies from classified industrial multi-modal knowledge, and develop an efficient and scalable tool for industrial anomaly detection.

\item We propose a novel LLM-Optimation module that evaluates and refines LLM-generated dependency graphs by validating nodes, edges, and overall logical consistency against industrial multi-modal knowledge. The module further supports feedback-driven graph optimation, ensuring accurate and meaningful dependency structures for anomaly detection.

\item We construct an Industrial Spatial-Temporal Graph (ISTG) that jointly models the dependencies among multi-sensor and actuator signals across both spatial and temporal dimensions. The multi-dimensional and multivariate anomaly detection substantially reduces false positives and false negatives.

\item Extensive experiments on 9 datasets demonstrate \sysname outperforms 12 SOTA methods in terms of both effectiveness and scalability. In addition, we built an industrial simulation platform and deployed a real-world robotic arm, where we implemented ICS attacks to collect datasets for further research in ICS security.

\end{itemize}

\noindent
\textbf{open-source: https://anonymous.4open.science/r/IstGPT-386A}

%% file: tex/Preliminaries.tex
\begin{figure*}[ht]
\centering
\includegraphics[width=\textwidth]{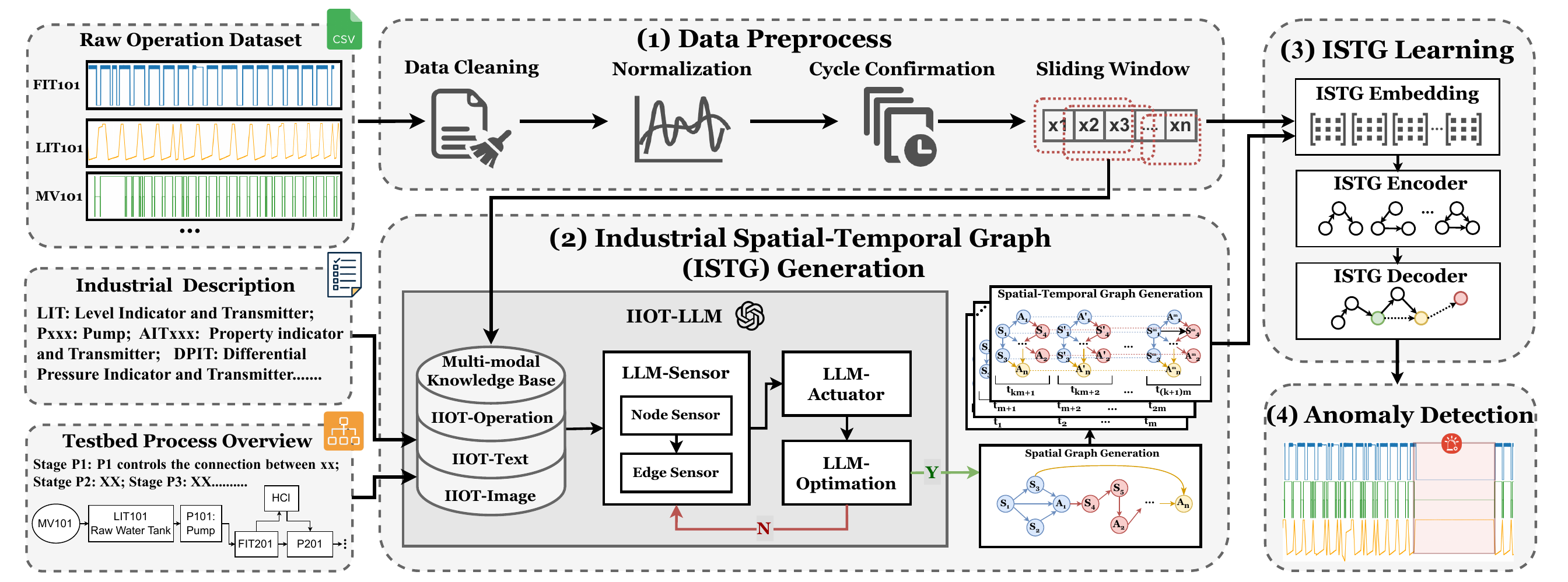}
\caption{The framework of \sysname.}
\label{fig_overview}
\end{figure*} 

\subsection{Preliminaries}

\textbf{ICS System.}
Industrial Control Systems (ICS) typically follow the Purdue Model Levels 0-2~\cite{holm2015survey}, covering control systems, intelligent devices, and physical processes \cite{conti2021survey}.
Purdue Level~2 hosts Supervisory Control and Data Acquisition (SCADA) systems responsible for supervisory control and monitoring~\cite{upadhyay2020scada}.
Purdue Level~1 includes intelligent devices such as Programmable Logic Controllers (PLCs), Remote Terminal Units (RTUs), and Intelligent Electronic Devices (IEDs), which execute control logic by reading sensor inputs and issuing actuator commands~\cite{wester2015role}.
Purdue Level~0 comprises physical sensors and actuators that directly interface with the industrial process~\cite{tomar2020plc}.

\textbf{Graph-based Industrial Anomaly Detection.}
Industrial anomaly detection aims to identify sensors or actuators exhibiting abnormal behavior over time. Formally, the $p$-th sensor or actuator is represented as a time series 
$\mathbf{sa}^p=[sa_1^p, sa_2^p, \ldots, sa_n^p]$, where $sa_i^p$ is the value at time step $i$.
Graph-based anomaly detection approaches model inter-variable dependencies using a graph $\mathcal{G}=(\mathcal{V},\mathcal{E})$, where nodes correspond to sensors or actuators and edges encode spatial or temporal relationships.

Graph-based methods commonly employ graph neural networks to reconstruct normal signal patterns and detect anomalies by comparing reconstructed values with ground truth. 
In particular, Graph Attention Networks (GATs)~\cite{velivckovic2017graph} assign attention weights $\alpha_{u,v}$ to neighboring nodes, enabling adaptive aggregation of relevant information:
{\small
\begin{equation}\label{eq_GAT_normal}
\alpha_{u,v}=
\frac{\exp\!\left(\mathrm{LeakyReLU}\!\left(\mathbf{a}_t^{\top}[\mathbf{W}\mathbf{h}_u \| \mathbf{W}\mathbf{h}_v]\right)\right)}
{\sum_{k\in\mathcal{N}(u)}
\exp\!\left(\mathrm{LeakyReLU}\!\left(\mathbf{a}_t^{\top}[\mathbf{W}\mathbf{h}_u \| \mathbf{W}\mathbf{h}_k]\right)\right)}
\end{equation}
}

$\mathbf{W}$ is a learnable linear transformation, $\mathbf{a}_t$ is an attention vector, and $\mathcal{N}(u)$ denotes the neighbors of node $u$, $(u,v,k) \in \mathcal{V}$.
For each $\mathbf{sa}^p$, anomalies are detected based on whether the condition $anomaly\_scores(sa^p_i, \hat{sa}^p_i)>threshold$ is satisfied, where $\hat{sa}^p_i$ is the value reconstructed from learned normal patterns.

\subsection{Threat Model}
ICS attacks span both the cyber and physical layers of the Purdue model. To the best of our knowledge, most ICS attacks ultimately target industrial sensors and actuators at Purdue Level~0~\cite{emake2020industrial, koay2023machine}. Our work focuses on detecting anomalous behaviors of sensors and actuators.

Attackers have compromised SCADA systems and deployed Industroyer malware to manipulate sensors and actuators through industrial protocols such as IEC-104 and IEC-61850~\cite{salazar2024tale}, to disrupt normal operations.
We consider two attack scenarios: (i) a remote attacker with access to the local plant communication network, and (ii) an on-site attacker with physical access to field devices. 
Attackers are assumed to possess general knowledge of industrial protocols and control architectures, but not detailed system-specific configurations. In both scenarios, the attacker can interact with a target PLC over industrial Ethernet (e.g., EtherNet/IP or Modbus/TCP). PLCs organize sensor and actuator data into register blocks for access. The attackers locate target register addresses via the PLC symbol table. They read the current values, offset them by a certain margin, and periodically write the modified values to the PLC output area (Area.PA) to maintain the offset. By communicating with PLCs, the malware issued unauthorized control commands, potentially leading to critical safety incidents.

%% file: tex/Approach.tex
We propose \sysname, a novel LLM-based anomaly detection framework for industrial cyber-physical systems, shown in Fig.~\ref{fig_overview}. \sysname leverages LLMs to construct a sensor-actuator spatial dependency graph and utilizes deep learning techniques to model industrial spatial-temporal patterns.

\sysname consists of four phases,
    \emph{Data Preprocess}, \emph{ISTG Generation}, \emph{ISTG Learning}, and \emph{Anomaly Detection}.
(1) \emph{The data preprocess phase:} enhancing the quality of data across multivariate sensors and actuators in industrial scenarios.
(2) \emph{The ISTG generation phase:} leveraging LLMs to generate a sensor-actuator dependency graph from classified industrial multi-modal knowledge, which is further extended into an Industrial Spatial-Temporal Graph (ISTG) to model spatial and temporal dependencies jointly.
(3) \emph{The ISTG learning phase:} embedding the ISTG and employing unsupervised deep learning to train an anomaly detection model. 
(4) \emph{The anomaly detection phase:} identifying and localizing anomalous sensors and actuators based on the anomaly detection model. The details of each phase will be introduced.

\subsection{Data Preprocess}
The data preprocessing includes three steps: 1) Data Cleaning, 2) Normalization, 3) Cycle Confirmation and Sliding Window. 


\subsubsection{Data Cleaning}
In complex industrial environments, raw SCADA logs inevitably contain \emph{dirty} data, such as measurement noise, missing values, and irrelevant records~\cite{lin2018tabor}. 
We therefore perform data cleaning before model training and inference. 
Given a sensor or actuator time series of length $n$, it is denoted as $\mathbf{{sa}} = [sa_1, sa_2, \ldots, sa_n] \in \mathbb{R}^{1 \times n} $. 
Missing values are handled via linear interpolation~\cite{huang2021missing}, which is suitable for industrial signals exhibiting locally linear behavior over short intervals~\cite{li2020relation, wan2023sensor}. 
To suppress high-frequency fluctuations, we apply a moving-average filter~\cite{macaulay1931introduction}, a widely adopted technique in prior ICS data-cleaning pipelines~\cite{lin2018tabor, golestan2013moving}. 

This step primarily mitigates routine noise that is short-lived and weakly structured. 
In contrast, attack-induced anomalies typically manifest as persistent or structured manipulations (e.g., sustained offsets or repeated tampering), which disrupt normal temporal patterns and cross-variable consistency. 
As a result, the anomalies detected by \sysname are more indicative of adversarial manipulations.

\subsubsection{Normalization}
To normalize data, we employ the Min-Max Normalization method~\cite{ali2022investigating}. 
The method has been widely used in deep learning to accelerate model training~\cite{huang2023normalization}. The preprocessed data from $p$ sensors and actuators with a length of n can be represented as $\mathbf{\tilde{SA}}=\left[\mathbf{\tilde{sa}^1}, \mathbf{\tilde{sa}^2}, \cdots, \mathbf{\tilde{sa}^p}\right] \in \mathbb{R}^{p \times n}$. Here, each preprocessed sensor or actuator is denoted by $\mathbf{\tilde{sa}}$.

\begin{algorithm}[t]
\caption{Industrial Cycle Estimation via Autocorrelation}
\label{algo:cycle_conf}
\SetKwInOut{Input}{Input}
\SetKwInOut{Output}{Output}
\Input{$\mathbf{\tilde{sa}}\in\mathbb{R}^{n}$: preprocessed multivariate sensor/actuator data;
$lag_{\min}, lag_{\max}$: search range;
$\tau$: peak threshold}
\Output{$T$: estimated industrial cycle (in samples)}
\tcp{(1) Compute autocorrelation function}
$ICS\_ACF \gets \emptyset$\; 
\For{$lag \gets lag_{\min}$ \KwTo $lag_{\max}$}{
    $ICS\_ACF[lag] \gets \mathrm{ICS\_ACFCaculator}(\tilde{\mathbf{sa}}, lag)$\;
}
\tcp{(2) Find significant peaks of $ICS\_ACF$}
$Peaks \gets \emptyset$\;
\For{$lag \gets lag_{\min}+1$ \KwTo $lag_{\max}-1$}{
    \If{$ICS\_ACF[lag] > ICS\_ACF[lag-1]$ \textbf{and} $ICS\_ACF[lag] > ICS\_ACF[lag+1]$ \textbf{and} $ICS\_ACF[lag] \ge \tau$}{
        Add($Peaks$, $lag$)\;
    }
}
\If{$IsEmpty(Peaks)$}{
    \Return $T \gets 0$\;
}
\tcp{(3) Choose the most significant peak as the cycle}
\Return $T \gets \arg\max_{lag \in Peaks} ICS\_ACF[lag]$\;
\end{algorithm}

\subsubsection{Cycle Confirmation and Sliding Window}
The sliding window technique segments time-series data into fixed-length portions for localized feature extraction, reducing computation and capturing temporal patterns.
To determine an appropriate window length, we first estimate the industrial process cycle.
The cycle estimation procedure is described in Algorithm~\ref{algo:cycle_conf}. 
The algorithm is based on autocorrelation, as defined: 
\begin{equation}\label{eq_slid_wind}
{\mathrm{ICS\_ACFCaculator}(\tilde{\mathbf{sa}}^p, lag)}=\frac{1}{n-\operatorname{lag}} \sum_{i=1}^{n-\operatorname{lag}} \tilde{sa}^p_{i} \cdot \tilde{sa}^p_{i+\operatorname{lag}}
\end{equation}
where $n$ is the length of the time series, $lag$ is the interval between two time points, and $\tilde{sa}_i$ denotes preprocessed data.

Given a window length $w$ and a stride $s$, a sliding window moves along the time axis to generate subsequences. The $i$-th segment is denoted as $\mathbf{d}_i=\left[\tilde{sa}_i, \tilde{sa}_{i+1}, \cdots, \tilde{sa}_{i+w-1}\right] \in \mathbb{R}^{1 \times w}, \; i \in (1+j \cdot s, j \in [0,wn-1])$. The total number of sliding windows $wn$ is calculated as follows:
\begin{equation}\label{eq_slid_wind2}
    wn=\left\lfloor\frac{n-w}{s}\right\rfloor+1, \; (w \geq T)
\end{equation}
$T$ denotes the estimated industrial cycle. Finally, the preprocessed data $\mathbf{SA}$ will be slid into $\mathcal{W} = [\mathbf{w^1}, \mathbf{w^2}, \dots, \mathbf{w^p}] \in \mathbb{R}^{p \times wn \times w}$, $\mathbf{w} = [\mathbf{d}_1, \mathbf{d_2}, \dots, \mathbf{d_{wn}}]\in \mathbb{R}^{wn \times w}$.

\begin{figure*}[ht]
\centering
\includegraphics[width=\linewidth]{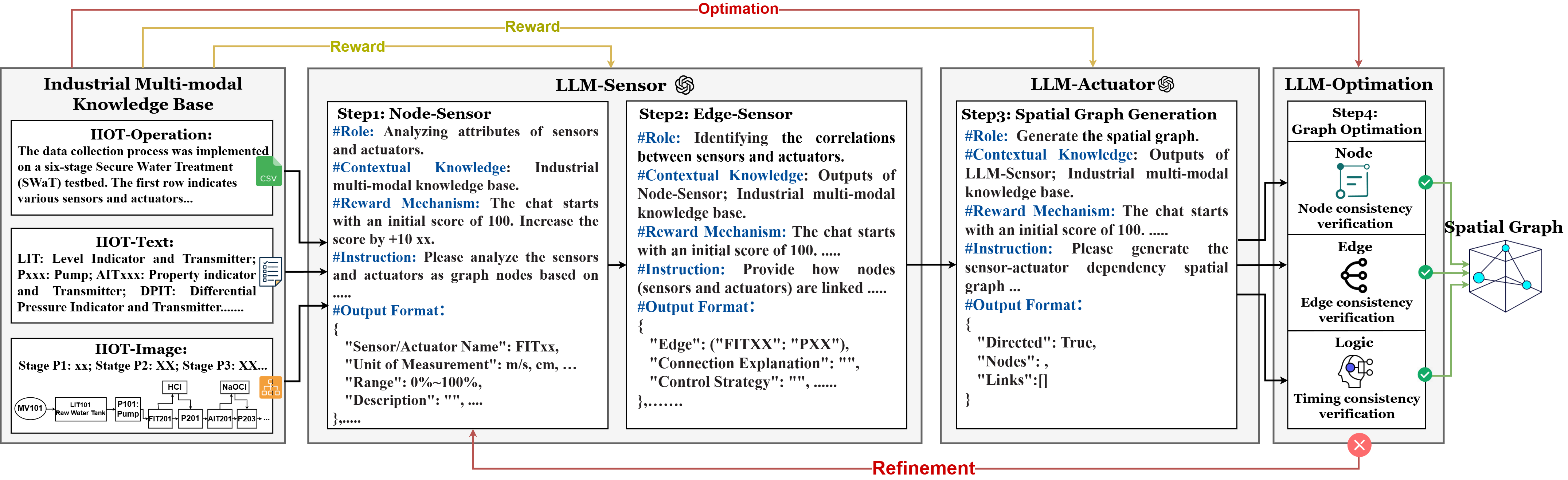}
\caption{The overall pipeline of \llm.}
\label{fig_graph_gen}
\end{figure*}

\subsection{LLM-based ISTG Generation}
\label{sec_graph_gen}

We proposed \llm, a multi-stage prompt engineering, as illustrated in Figure~\ref{fig_graph_gen}. 
1) We classify an industrial multi-modal knowledge base. 2) The prompt-engineering pipeline consists of LLM-Sensor, LLM-Actuator, and LLM-Optimation, and iteratively guides graph generation and optimization. 3) We construct the Industrial Spatial-Temporal Graph (ISTG).

\subsubsection{Industrial multi-modal Knowledge Base}
In industrial Cyber-Physical Systems (CPS), constructing accurate variable dependency graphs is challenging due to complex system behaviors and the high dimensionality induced by hundreds of sensors and actuators~\cite{zhai2021structured}. 
Existing graph learning-based approaches~\cite{li2023staged, zhan2022stgat} that rely solely on sensor and actuator measurements are often insufficient. 
To address this limitation, incorporating industrial prior knowledge, such as technical documents and instrumentation diagrams, is essential~\cite{karniadakis2021physics, wu2022data}. 
Accordingly, we summarize and organize industrial multimodal knowledge to support LLM-based dependency graph generation.
The industrial multi-modal knowledge base is composed of IIOT-Operation, IIOT-Text, and IIOT-Image, defined as Eq.\ref{eq_knowledge}. Details are as follows:
\begin{equation}\label{eq_knowledge}
    \mathcal{K}=\{\mathcal{K}^{\mathrm{op}},\,\mathcal{K}^{\mathrm{txt}},\,\mathcal{K}^{\mathrm{img}}\}
\end{equation}

\textbf{IIOT-Operation} refers to the operational data of sensors and actuators over specific periods in the Industrial Internet of Things (IIOT). The operational data typically includes temperature, pressure, flow rate, and other parameters essential for SCADA systems. 
IIOT-Operation captures the data-driven dependencies, enabling the LLM to understand sensor-actuator interactions and temporal dynamics.

\textbf{IIOT-Text} refers to textual descriptions of industrial systems, such as sensor and actuator specifications, standard operating procedures, and data collection workflows.
By incorporating textual knowledge, the LLM gains explicit semantic context, enabling it to reason more effectively about variable roles and dependencies.

\textbf{IIOT-Image} refers to visual information of industrial processes, including testbed workflows, system layouts, and device interactions depicted in diagrams or schematics.
Such visual information assists the LLM in identifying patterns of physical interaction and spatial relationships, contributing to a graph structure more aligned with the real-world system. 
However, industrial diagrams (e.g., P\&IDs) typically expose only partial dependencies~\cite{zhai2021structured}, making IIOT-Image alone insufficient for complete graph construction. 
Therefore, IIOT-Image is jointly used with IIOT-Operation and IIOT-Text.

Given the industrial multi-modal knowledge base $\mathcal{K}$, IIOT-LLM (Eq.\ref{eq_IIOTLLM}) guides the LLM to construct a comprehensive and interpretable spatial dependency graph. 
Specifically, a Retrieval Augmented Generation (RAG) module $\mathcal{R}(\cdot)$ retrieves task-relevant context from contextual knowledge $\mathcal{CK}$, which is combined with the task instruction via the prompt construction function $\Pi(\cdot)$. 
The prompt is then processed by the adopted LLM $\mathcal{M}_{\theta}$, where $\theta$ denotes the model parameters and configuration.
\begin{equation}\label{eq_IIOTLLM}
    \textsc{IIOT\text{-}LLM}(task;\{\mathcal{CK}\})=\mathcal{M}_{\theta}\!\left(\Pi(task,\mathcal{R}(\{\mathcal{CK}\},task))\right)   
\end{equation}

Notably, the RAG-driven context selection restricts the prompt to a bounded set of task-relevant fragments rather than the full knowledge base. Therefore, \llm effectively limits the prompt size within the practical context capacity of current LLMs, even for large-scale systems with thousands of sensors and actuators.

\subsubsection{Prompt Engineering}
Prompts are structured token sequences that guide LLMs to follow task-specific rules and produce high-quality outputs~\cite{liu2023prompting, chen2023unleashing}.
A naive approach to dependency graph generation is to directly prompt an LLM with industrial multimodal knowledge (e.g., "\emph{Generate a spatial graph based on the following domain knowledge}").
However, such monolithic prompts limit step-by-step reasoning and task decomposition, often resulting in inconsistent or low-quality graphs.
To address this limitation, we design a multi-stage zero-shot prompt engineering pipeline that decomposes graph generation into three modules: \emph{LLM-Sensor}, \emph{LLM-Actuator}, and \emph{LLM-Optimation}.

Firstly, LLM-Sensor is composed of \emph{Node-Sensor} and \emph{Edge-Sensor}, which generate nodes and edges, respectively.
Node-Sensor analyzes sensor and actuator attributes (e.g., names, units, and ranges) to produce structured node representations.
Formally, given a node-analysis $task_{node}$ and the industrial knowledge base $\mathcal{K}$, Node-Sensor produces textual node descriptions, which are transformed into a structured node set $\hat{\mathcal{V}}_{NS}$ by the parsing operator $\mathcal{P}_{V}$:
\begin{equation}\label{eq_node-sensor}
\hat{\mathcal{V}}_{NS}=\mathcal{P}_{V}\!\left(\textsc{IIOT\text{-}LLM}(task_{node};\{\mathcal{K}\})\right),
\end{equation}
where each node $\hat{v}_i^{S}=(sa_i,\hat{a}_i,\hat{t}_i)\in\hat{\mathcal{V}}_{NS}$ corresponds to a sensor or actuator variable $sa_i$ with inferred attributes $\hat{a}_i$ and textual descriptions $\hat{t}_i$.

Edge-Sensor infers directed dependencies $\hat{\mathcal{E}}_{ES}$ among nodes in $\hat{\mathcal{V}}_{NS}$ using the knowledge base $\mathcal{K}$ (Eq.\ref{eq_edge-sensor}).
Each inferred edge $\hat{e}_{ij}^{S}=\left((\hat{v}_i^{S},\hat{v}_j^{S}),\,\hat{r}_i\right)\in{\hat{\mathcal{E}}}_{ES}$ represents a dependency from $\hat{v}_i^{S}$ to $\hat{v}_j^{S}$ with an associated explanation $\hat{r}_{ij}$.
The raw textual outputs generated by the LLM are transformed into the structured edge set via the parsing operator $\mathcal{P}_{E}$:
\begin{equation}\label{eq_edge-sensor}
\hat{\mathcal{E}}_{ES}=\mathcal{P}_{E}\!\left(\textsc{IIOT\text{-}LLM}(task_{edge};\{\mathcal{K},\hat{\mathcal{V}}_{NS}\})\right).
\end{equation}

The outputs of Node-Sensor $\hat{\mathcal{V}}_{NS}$ and Edge-Sensor $\hat{\mathcal{E}}_{ES}$ are integrated by LLM-Actuator to construct an initial sensor-actuator dependency graph $\hat{\mathcal{G}}_{LA}$. Specifically, LLM-Actuator generates a graph-level textual description based on the inferred nodes and edges, which is transformed into a structured graph representation via the parsing operator $\mathcal{P}_{G}$:
\begin{equation}\label{eq_LLM-Actuator}
\hat{\mathcal{G}}_{LA}=\mathcal{P}_{G}\!\left(\textsc{IIOT\text{-}LLM}(task_{graph};\{\mathcal{K},\hat{\mathcal{V}}_{NS},\hat{\mathcal{E}}_{ES}\})\right).
\end{equation}

To ensure consistency with industrial knowledge $\mathcal{K}$, the generated graph is iteratively validated and refined by LLM-Optimation across three constraints: node validity ($\mathcal{C}^{node}$), edge consistency ($\mathcal{C}^{edge}$), and logical coherence ($\mathcal{C}^{logic}$):
\begin{equation}\label{eq_LLM-optimation}
    \mathcal{C}=\{\mathcal{C}^{node},\mathcal{C}^{edge},\mathcal{C}^{logic}\}, \mathcal{C}^* = \mathcal{R}(\mathcal{K},*) \subseteq \mathcal{K}
\end{equation}
The above constraints are not manually defined rules. Instead, they are automatically derived via the retrieval function $\mathcal{R}(\cdot)$ from the heterogeneous industrial knowledge base $\mathcal{K}$. 
Moreover, $\mathcal{C}^{node}$, $\mathcal{C}^{edge}$, and $\mathcal{C}^{logic}$ depend on the intermediate graph structure produced by LLM-Sensor and LLM-Actuator, and therefore cannot be fully specified a priori within the knowledge base. This necessitates an iterative validation-and-refinement process to progressively correct structural inconsistencies and improve graph reliability.

If any constraint is violated, LLM-Optimation generates structured feedback ($\Delta_V, \Delta_E, \Delta_L$) that is appended to subsequent prompts, triggering regeneration by LLM-Sensor and LLM-Actuator.
This iterative process continues until all constraints are satisfied or the maximum iteration count $K_{\max}$ is reached, yielding a refined spatial graph $\mathcal{G}_{spatial}$.
The overall optimization procedure is summarized in Algorithm~\ref{alg:iiot_optimization}.

\begin{algorithm}[tbp]
\caption{IIOT-Optimization: Node/Edge/Logic-Guided Graph Validation and Refinement}
\label{alg:iiot_optimization}

\SetKwInOut{Input}{Input}
\SetKwInOut{Output}{Output}

\Input{Initial spatial graph $\hat{\mathcal{G}}^{(0)}_{LA}$; industrial multi-modal knowledge base $\mathcal{K}$; domain constraints $\mathcal{C}$; maximum iterations $K_{\max}$}
\Output{Validated and refined graph $\mathcal{G}_{spatial}$}

\tcp{Notation: $\oplus$ denotes appending feedback to a prompt}
\tcp{Initialization}
$k \gets 0;\ \Delta_V, \Delta_E, \Delta_L \gets \emptyset$\;

\While{$k < K_{\max}$}{
    \tcp{(1) Node-optimization}
    $\Delta_V \gets \textsc{IIOT-LLM}\!\bigl(task_{checknodes(\hat{\mathcal{G}}^{(k)}_{LA})};
    \{\mathcal{C}^{node}\}\bigr)$\;

    \tcp{(2) Edge-optimization}
    $\Delta_E \gets \textsc{IIOT-LLM}\!\bigl(
    task_{checkedges(\hat{\mathcal{G}}^{(k)}_{LA})};
    \{\mathcal{C}^{edge}\}\bigr)$\;

    \tcp{(3) Logic-optimization}
    $\Delta_L \gets \textsc{IIOT-LLM}\!\bigl(
    task_{checklogic(\hat{\mathcal{G}}^{(k)}_{LA})};
    \{\mathcal{C}^{logic}\}\bigr)$\;

    \If{$\Delta_V=\emptyset \; \textbf{and} \Delta_E=\emptyset \; \textbf{and} \Delta_L=\emptyset$}{
        \Return $\mathcal{G}_{spatial} \gets \hat{\mathcal{G}}^{(k)}_{LA}$\;
    }

    \tcp{(4) Graph refinement}
    $task'_{node} \gets task_{node} \oplus \Delta_V \oplus \Delta_L$\;

    $\hat{\mathcal{V}}^{(k+1)}_{NS} \gets
    \mathcal{P}_V\!\bigl(
    \textsc{IIOT-LLM}(task'_{node};\{\mathcal{K}\})\bigr)$\;

    $task'_{edge} \gets task_{edge} \oplus \Delta_E \oplus \Delta_L$\;

    $\hat{\mathcal{E}}^{(k+1)}_{ES} \gets
    \mathcal{P}_E\!(
    \textsc{IIOT-LLM}(task'_{edge};
    \{\mathcal{K}, \hat{\mathcal{V}}^{(k+1)}_{NS}\}))$\;

    $task'_{graph} \gets task_{graph} \oplus \Delta_L$\;

    $\hat{\mathcal{G}}^{(k+1)}_{LA} \gets
    \mathcal{P}_G\!\bigl(
    \textsc{IIOT-LLM}(task'_{graph};
    \{\mathcal{K}, \hat{\mathcal{V}}^{(k+1)}_{NS},
    \hat{\mathcal{E}}^{(k+1)}_{ES}\})\bigr)$\;

    \tcp{(5) Iteration}
    $k \gets k + 1$\;
}
\Return $\mathcal{G}_{spatial} \gets \hat{\mathcal{G}}^{(k)}_{LA}$\;

\end{algorithm}

Each module adopts a unified prompt template consisting of five components: \emph{Role}, \emph{Contextual Knowledge}, \emph{Reward Mechanism}, \emph{Instruction}, and \emph{Output Format} (Table~\ref{tab:tab_prompt}).
The \emph{Role} constrains the LLM’s responsibility. The \emph{Contextual Knowledge} ensures outputs align with the specific domain and scenario needs.
The \emph{Reward Mechanism} enhances its tendency to generate high-quality content. \emph{Instruction} defines the exact task. And \emph{Output Format} enforces structured, parseable outputs for downstream processing.
Prompt patterns for each module, organized by these components, are provided in Table~\ref{tab:prompt_patterns}.

\input{table/Approach/prompt}
\input{table/Approach/details_prompts}

\begin{figure}[tbp]
\centering
\includegraphics[width=0.4\textwidth]{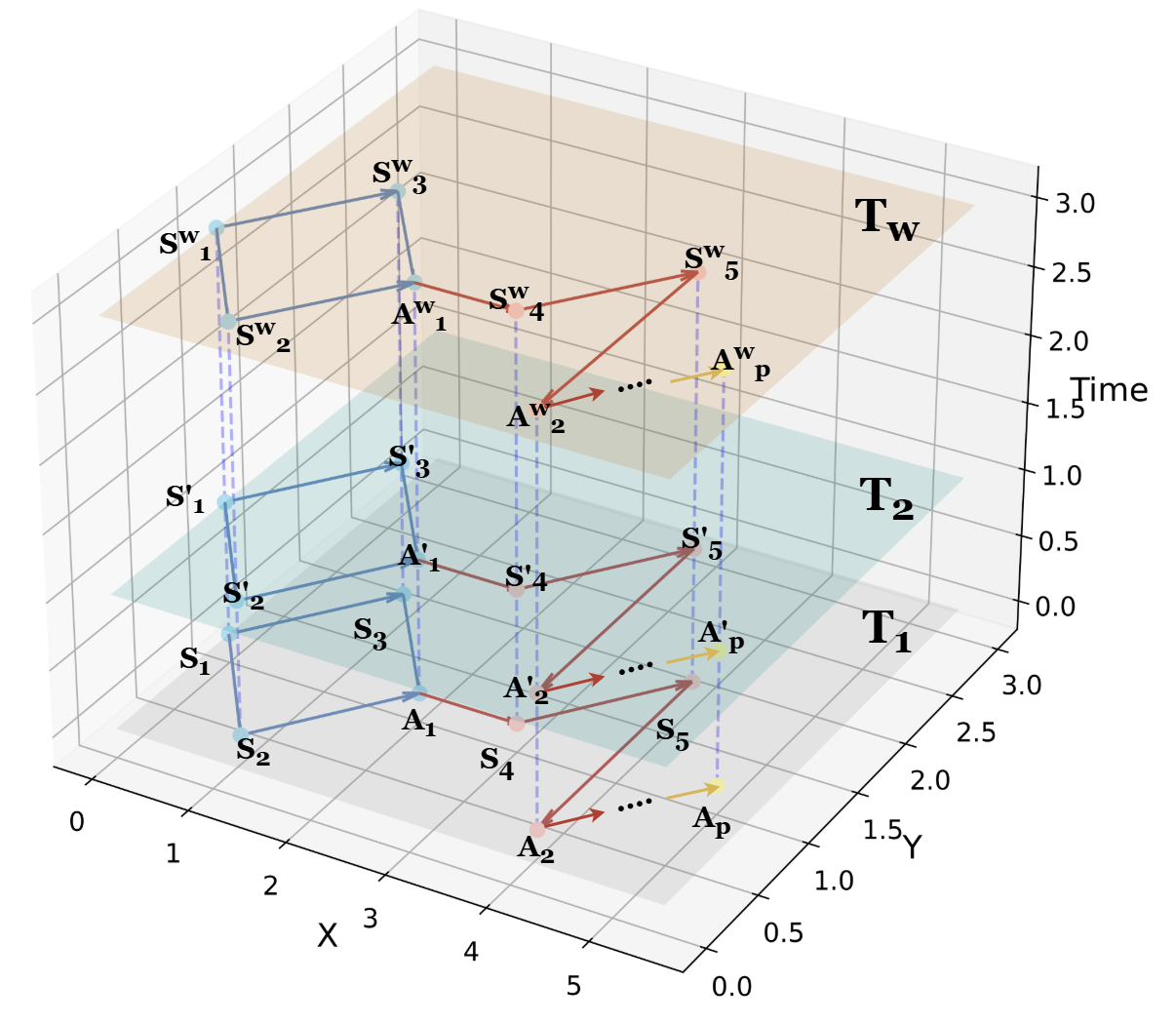}
\caption{\label{fig_st-graph} The ISTG in each sliding window: colored solid arrows indicate spatial edges and colored dashed lines indicate temporal edges. \textbf{$S^y_x$} indicates $sensor_x$ at time $y$ and $A^y_x$ indicates $actuator_x$ at time $y$.}
\end{figure}

\subsubsection{ISTG Construction}
Based on the spatial dependency graph $\mathcal{G}_{spatial}$, \sysname incorporates temporal correlations to construct the Industrial Spatial-Temporal Graph (ISTG), as illustrated in Fig.~\ref{fig_st-graph}.

In the spatial dimension, the preprocessed sensor and actuator data are modeled as a directed graph
$\mathcal{G}_{spatial}=(\mathcal{V}^s,\mathcal{E}^s)$, where $\mathcal{V}^s$ denotes the set of $p$ sensor/actuator nodes and $\mathcal{E}^s$ denotes the spatial dependency edges:
{\small
\begin{equation}\label{eq_spatial_graph}
\mathcal{V}^s=\{v_j \mid j\in[1,p]\},\quad
\mathcal{E}^s=\{e^s_{i,j}\mid (i,j)\in\mathcal{G}_{spatial}[src,dst]\}.
\end{equation}
}
At time step $i$, each node is represented as
$ v_j^{(i)} = (label_{j}, value_{j}^{(i)}, i), j \in [1,p], i \in [1,n]$,
where $label_j$ specifies the variable identity and type (sensor or actuator), and $value_j^{(i)}$ denotes the measurement at time $i$.
Each spatial edge is defined as
$e^s_{i,j}=(v_i,v_j,0)$,
where type $0$ indicates a spatial dependency between variables.

To capture temporal correlations, we introduce directed temporal edges that connect the same variable across consecutive time steps within each sliding window.
The nodes within each sliding window are denoted by $\mathcal{V}$, and the temporal edge set is denoted by $\mathcal{E}^t$, as defined in equation~\ref{eq_ISTG}. 
{\small
\begin{equation}\label{eq_ISTG}
\begin{aligned}
&\mathcal{V} = \{v \in \mathcal{V}^{s}_{i}|i \in [k,k+w-1], k \in (1+j \cdot s, j \in [0,wn-1])\} \\
&\mathcal{E}^t = \{e^t_{i,i+1}|i \in [k,k+w-1], k \in (1+j \cdot s, j \in [0,wn-1])\}
\end{aligned}
\end{equation} 
}
$\mathcal{V}^{s}_{i}$ indicates the node set at time $i$. Specifically, each edge is represented as $e^t_{i,i+1}=(v_j^i,v_j^{i+1},1) \in \mathcal{E}^t, i \in [1,n-1], j \in [1,p]$, where $e^t_{i,i+1}$ indicates a directed edge from node $v_j$ at time $i$ to node $v_j$ at time $i+1$ with type 1. Type 1 indicates temporal connections.

Finally, \sysname creates the ISTG, which is defined as $\mathcal{G} = (\mathcal{V}, \mathcal{E}^s, \mathcal{E}^t)$.

\subsection{ISTG Learning} 
\begin{figure*}[tbp]
\centering
\includegraphics[width=0.9\textwidth]{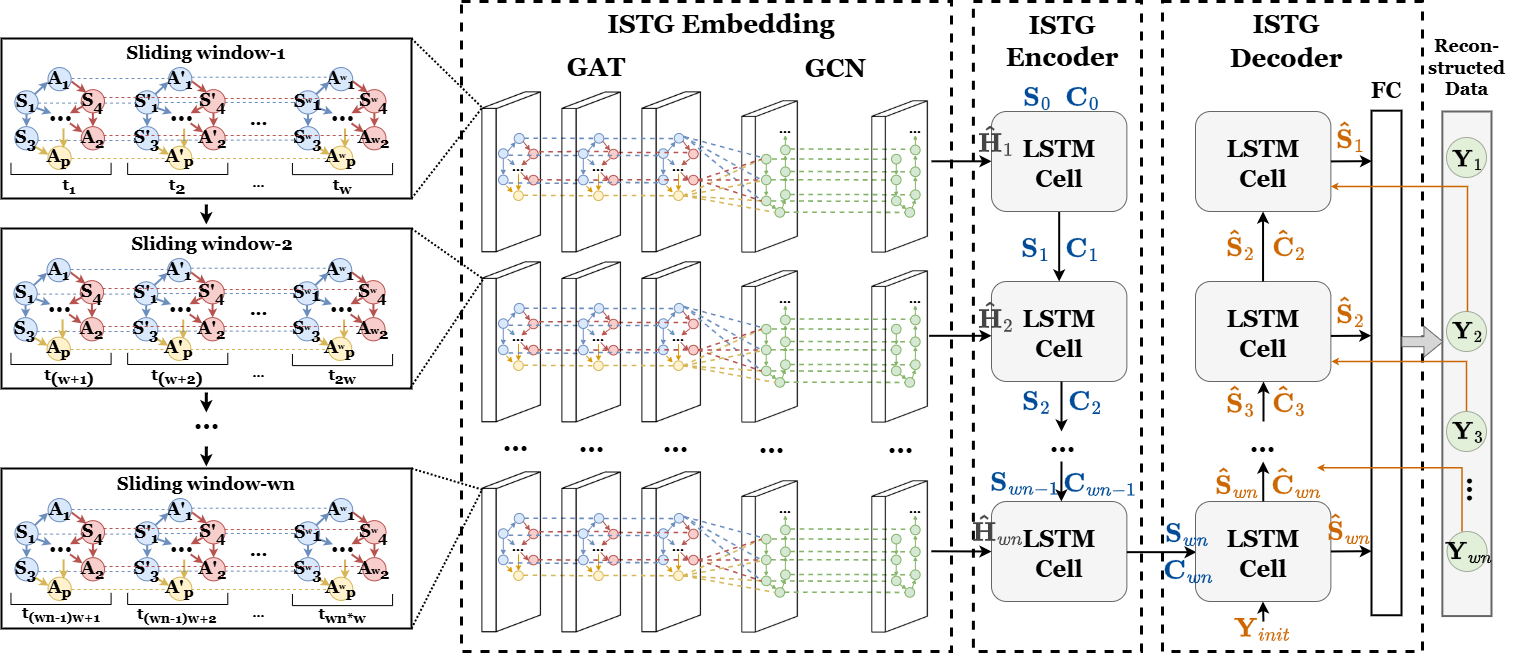}
\caption{The workflow of ISTG Learning.}
\label{fig_graph_learning}
\end{figure*}

\sysname performs unsupervised learning on the ISTG to model normal spatial-temporal behaviors. As illustrated in Fig.~\ref{fig_graph_learning}, the learning pipeline consists of three stages:
1) ISTG embedding, 2) temporal encoding, and 3) data reconstruction.

\subsubsection{ISTG Embedding} 
For each length of w sliding window, ISTG first embeds nodes into vectors $\mathbf{H} \in \mathbb{R}^{p \times w}$, where $p$ is the total number of sensors and actuators. The node vectors are denoted by $\mathcal{H} \in \mathbb{R}^{wn \times p \times w}$ as follows:

\begin{equation}\label{eq_node_embedding}
    \mathcal{H}={[\mathbf{H}_1,\mathbf{H}_2,...,\mathbf{H}_{wn}]},
    \mathbf{H}_t={[\mathbf{h}_{(t,1)},\mathbf{h}_{(t,2)},...,\mathbf{h}_{(t,p)}]}
\end{equation} 

For each sensor or actuator, $\mathbf{h}_{(t,i)} \in \mathbb{R}^w$ denotes the feature vector of node $i$ in the $t$-th sliding window. Then, the node embeddings are interconnected through edges of various types. The spatial edge set $\mathcal{E}^s$ is vectorized as $\mathcal{L}^s \in \mathbb{R}^{2\times|\mathcal{E}^s|}$ and the temporal edge set $\mathcal{E}^t$ is vectorized as $\mathcal{L}^t \in \mathbb{R}^{2\times|\mathcal{E}^t|}$.

\sysname then take vectorized ISTG $\mathcal{G}=(\mathcal{H}, \mathcal{L}^s,\mathcal{L}^t)$ as inputs for message passing, which involves aggregation and update steps. In the aggregation step, each node aggregates features from its neighboring nodes in the previous layer. In the update step, each node updates its representation using the aggregated information. 

\sysname employs graph attention convolution (GATConv)~\cite{velivckovic2017graph} for spatial message passing and graph convolutional network (GCNConv)~\cite{zhang2019graph} for temporal propagation. 
We use GATConv in the spatial domain because the dependency edges among sensors and actuators exhibit heterogeneous strengths. Higher data interaction indicates stronger coupling, while lower interaction implies weaker relationships. 
The attention mechanism in GATConv naturally captures such variability. 
In contrast, temporal edges of each variable are treated with equal importance. Therefore, the lightweight and efficient GCNConv is sufficient for temporal modeling without introducing unnecessary attention overhead.

Besides, we do not adopt the latest graph learning architectures such as Graph Transformers~\cite{shehzad2026graph}, because their global self-attention mechanism incurs $O(N^2)$ time and memory complexity. In large-scale industrial control systems with thousands of sensors and actuators, such quadratic overhead poses a significant barrier to practical deployment.

In GATConv layer, \sysname use $(\mathcal{H},\mathcal{L}^s)$ as inputs. GATConv leverages an attention mechanism to assign weights $\alpha_{u,v}$ to neighbor nodes and aggregate the most relevant node information. The attention coefficients $a_{u,v}$ in the $t$-th sliding window are computed as in equation~\ref{eq_GAT_normal}. 
The coefficient $a_{u,v}$ determines how much attention node $u \in \mathcal{L}^s[src]$ should pay to node $v \in \mathcal{L}^s[dst]$ when aggregating information. After aggregating the information of the neighbor nodes in the previous layer $l$, the node features $h_{(t,v)}^{(l+1)}$ are updated using equation \ref{eq_GAT_trans}. $\sigma (\cdot)$ is activation function sigmoid.

{\small
\begin{equation}\label{eq_GAT_trans}
    \mathbf{h}_{(t,v)}^{(l+1)}=\sigma\left(\sum_{u \in \mathcal{N}(v)} \alpha_{u,v}^{(l)} \mathbf{W}^s_t \mathbf{h}_{(t,u)}^{(l)}\right)
\end{equation}
}

In the GCNConv layer, \sysname takes $(\mathbf{H}_{GAT}, \mathcal{L}^t)$ as inputs, where $\mathbf{H}_{GAT}$ denotes the feature representations produced by the GATConv layer after nonlinear activation and tensor reshaping. 
In time series, the importance of adjacent time steps is uniform. Thus, the equal-weighted aggregation of GCN is well-suited. The node $v \in \mathcal{L}^t[dst]$ aggregates the features of the neighbor node $u \in \mathcal{L}^t[src]$ in the previous layer $r$ and the embedding of the node $\mathbf{\hat{h}}_{(k,v)}^{(r+1)}$ in the sliding window $k$ will be updated by equation~\ref{eq_GCN_trans}. $\tilde{A}$ is the adjacency matrix (with self-loops) for temporal propagation, and $\tilde{D}$ is its degree matrix.

\begin{equation}\label{eq_GCN_trans}
\begin{aligned}
&\mathbf{\hat{h}}^{(r+1)}_{(k,v)}
=
\sigma\!\left(
\sum_{u \in \mathcal{N}(v)\cup\{v\}}
\tilde{D}^{-\frac12}\tilde{A}\tilde{D}^{-\frac12}[v,u]\;
\mathbf{W}^{t}_{k}\,\mathbf{\hat{h}}^{(r)}_{(k,u)}
\right) 
\end{aligned}
\end{equation}
After message passing, the ISTG embeddings are further encoded to model inter-window representations and global spatio-temporal dependencies.

\subsubsection{ISTG Encoder}
The ISTG encoder uses long-short-term memory (LSTM) cells~\cite{sherstinsky2020fundamentals} to capture temporal dependencies across sliding windows. The cell state $C^{(t)}$ through three gates: input gate($i_t$), forget gate($f_t$), and output gate($o_t$) effectively captures the long-term dependencies in the ISTG. \sysname use the graph embeddings $\mathbf{\hat{H}}$ as input. In the $t$-th sliding window, the node $v$ hidden state $s^{(t)}_v$ will be updated by equation~\ref{eq_lstm_encoder}. 
\begin{equation}\label{eq_lstm_encoder}
\begin{aligned}
&\mathbf{i}_{(t,v)} =\sigma\left(\mathbf{W}^i_{(t,v)}\begin{bmatrix}\mathbf{\hat{h}}_{(t,v)},\mathbf{s}^{(t-1)}_v\end{bmatrix}+\mathbf{b}^i_{(t,v)}\right) \\
&\mathbf{f}_{(t,v)}=\sigma\left(\mathbf{W}^f_{(t,v)}\begin{bmatrix}\mathbf{\hat{h}}_{(t,v)},\mathbf{s}^{(t-1)}_v\end{bmatrix}+\mathbf{b}^f_{(t,v)}\right) \\
&\mathbf{o}_{(t,v)} =\sigma\left(\mathbf{W}^o_{(t,v)}\begin{bmatrix}\mathbf{\hat{h}}_{(t,v)},\mathbf{s}^{(t-1)}_v\end{bmatrix}+\mathbf{b}^o_{(t,v)}\right) \\
&\mathbf{c}^{(t)}_v =\mathbf{f}_{(t,v)}\odot\mathbf{c}^{(t-1)}_v+\mathbf{i}_{(t,v)}\odot \tanh\Big( \mathbf{W}^c_{(t,v)}\begin{bmatrix}\mathbf{\hat{h}}_{(t,v)},\mathbf{s}^{(t-1)}_v\end{bmatrix}\\
&\quad +\mathbf{b}^c_{(t,v)} \Big) \\
&\mathbf{s}^{(t)}_v =\mathbf{o}_{(t,v)}\odot\tanh(\mathbf{c}^{(t)}_v) \\
\end{aligned}
\end{equation}
$\odot$ is element-wise multiplication between two matrices. $tanh$ is the activation function. The module encodes ISTG embeddings into a fixed-length latent representation for the ISTG decoder.

\subsubsection{ISTG Decoder}
ISTG decoder utilizes the last hidden state \(\mathbf{S}^{(wn)}\) and cell state \(\mathbf{C}^{(wn)}\) from ISTG encoder to reconstruct the window of sensors and actuators data, denoted as \( \mathbf{Y}_{out} = [\mathbf{Y}_1,\mathbf{Y}_2,...,\mathbf{Y}_{wn}] \in \mathbb{R}^{wn \times p \times w}, \mathbf{Y}_i=[\mathbf{y}^{(i)}_1,\mathbf{y}^{(i)}_2,...,\mathbf{y}^{(i)}_p] \). 
In the $t$-th sliding window, the node v reconstructed data $\mathbf{y}^{(t)}_v \in \mathbb{R}^w$ is calculated by equation~\ref{eq_lstm_decoder}. 
\begin{equation}\label{eq_lstm_decoder}
\allowdisplaybreaks
\begin{aligned}
&\mathbf{\hat{i}}_{(t,v)} =\sigma\left(\mathbf{\hat{W}}^i_{(t,v)}\begin{bmatrix}\mathbf{{y}}^{(t+1)}_v,\mathbf{\hat{s}}^{(t+1)}_v\end{bmatrix}+\mathbf{\hat{b}}^i_{(t,v)}\right) \\
&\mathbf{\hat{f}}_{(t,v)}=\sigma\left(\mathbf{\hat{W}}^f_{(t,v)}\begin{bmatrix}\mathbf{{y}}^{(t+1)}_v,\mathbf{\hat{s}}^{(t+1)}_v\end{bmatrix}+\mathbf{\hat{b}}^f_{(t,v)}\right) \\
&\mathbf{\hat{o}}_{(t,v)} =\sigma\left(\mathbf{\hat{W}}^o_{(t,v)}\begin{bmatrix}\mathbf{{y}}^{(t+1)}_v,\mathbf{\hat{s}}^{(t+1)}_v\end{bmatrix}+\mathbf{\hat{b}}^o_{(t,v)}\right) \\
&\mathbf{\hat{c}}^{(t)}_v =\mathbf{\hat{f}}_{(t,v)}\odot\mathbf{\hat{c}}^{(t+1)}_v+\mathbf{\hat{i}}_{(t,v)}\odot \tanh(\mathbf{\hat{W}}^c_{(t,v)}\begin{bmatrix}\mathbf{y}^{(t+1)}_v,\mathbf{\hat{s}}^{(t+1)}_v\end{bmatrix}\\  
& + \mathbf{\hat{b}}^c_{(t,v)}) \\
&\mathbf{\hat{s}}^{(t)}_v =\mathbf{\hat{o}}_{(t,v)}\odot\tanh(\mathbf{\hat{c}}^{(t)}_v) \\
&\mathbf{y}^{(t)}_v = FC(\mathbf{\hat{s}}^{(t)}_v) \\
\end{aligned}
\end{equation}

The computation of three gates (\(\hat{i}_t\), \(\hat{f}_t\), \(\hat{o}_t\)), the hidden state $\mathbf{\hat{s}}^{(t)}_v$, and the cell state $\mathbf{\hat{c}}^{(t)}_v$ remain the same as ISTG encoder. $FC(\cdot)$ denotes a fully connected layer outputting the reconstruction.
Unlike ISTG encoder, the decoder uses the previous reconstructed output \(\mathbf{y}^{(t+1)}_v\) as input for subsequent predictions. The initial input $\mathbf{Y}_{init}$ is usually a specific initialization value (i.e., a zero vector). 

Besides, the reconstructed loss function adopts methods of mean squared error (MSE)~\cite{jadon2024comprehensive} and Entropy~\cite{mao2023cross} to minimize the reconstructed error. The loss function $\mathcal{L}_{loss}$ of each node in the $t$-th sliding window will be expressed as equation~\ref{eq_loss}. $||\cdot||_2$ indicates the Euclidean distance between two vectors. $\mathbf{h}_{(t,v)}[j]$ represents the node embedding $v$ in the $t$-th sliding window at time $j \in [1,w]$. $\mathbf{y}^{(t)}_v[j]$ represents the reconstructed embedding $v$ in the $t$-th sliding window at time $j$.
{\small
\begin{equation}\label{eq_loss}
\begin{aligned}&\mathcal{L}_{loss}={\frac{1}{w}||\mathbf{y}^{(t)}_v-\mathbf{h}_{(t,v)}||^2_2}+\lambda\sum_{j=1}^wp(\mathbf{h}_{(t,v)}[j])\log\left(\frac{p(\mathbf{h}_{(t,v)}[j])}{q(\mathbf{y}^{(t)}_v[j])}\right)\\
&p(\mathbf{h}_{(t,v)}[j])=\frac{\exp(\mathbf{h}_{(t,v)}[j])}{\sum_{i=1}^w\exp(\mathbf{h}_{(t,v)}[i])},q(\mathbf{y}^{(t)}_v[j])=\frac{\exp(\mathbf{y}^{(t)}_v[j])}{\sum_{i=1}^w\exp(\mathbf{y}^{(t)}_v[i])}
\end{aligned}
\end{equation}
}

After ISTG encoder-decoder architecture processes ISTG embeddings, \sysname performs anomaly detection based on the reconstructed data.

\begin{figure}[htbp]
\centering
\includegraphics[width=0.55\textwidth]{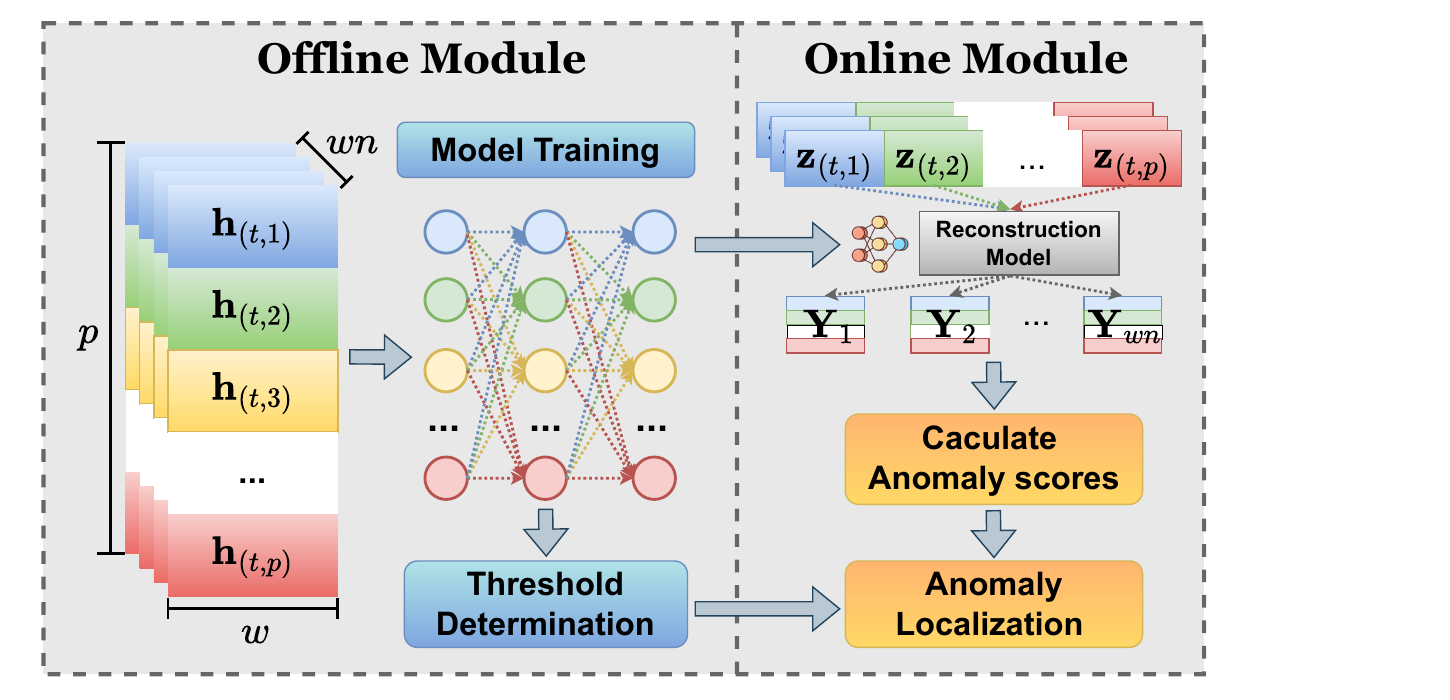}
\caption{\label{fig_detect} The implementation of Anomaly Detection.}
\end{figure}
\subsection{Anomaly Detection}
The anomaly detection phase consists of two modules: an offline module and an online module, as illustrated in Figure~\ref{fig_detect}.
During ISTG learning, \sysname trains the reconstruction model using normal data and determines an anomaly threshold, which is subsequently used for online detection.

In the online module, the preprocessed and embedded testing data $\mathbf{z}_{(t,1)}, \mathbf{z}_{(t,2)}, \ldots, \mathbf{z}_{(t,p)}$ are reconstructed into $\mathbf{y}^{(t)}_1, \mathbf{y}^{(t)}_2, \ldots, \mathbf{y}^{(t)}_p \in \mathbf{Y}_t$. 
Anomaly scores are computed based on the deviation between the reconstructed outputs and the ground truth, and anomalies are flagged by comparing the scores against the learned threshold.

\noindent
\textbf{Anomaly Scores.}
First, for each node $v$ in the $t$-th sliding window, we compute the L2 distance $L2\_dist_{(t,v)}$, cosine distance $cos\_dist_{(t,v)}$, and KL divergence $KL_{(t,v)}$, as defined in Eq.\ref{eq_ascore}.
These metrics are then normalized using an Interquartile Range (IQR)-based robust normalization (Eq.\ref{eq_normal}), where the median is subtracted and the result is divided by the corresponding IQR, producing three normalized scores: ${score}^{(L2)}{t,v}$, ${score}^{(cos)}{t,v}$, and ${score}^{(KL)}{t,v}$.
Here, $\varepsilon$ is a small positive constant added to avoid division by zero.
Finally, the overall anomaly score $\text{F}_\text{score}{(t,v)}$ is computed according to Eq.\ref{eq_fin_score}, where $\alpha$, $\beta$, and $\gamma$ are the hyperparameters to balance the weight of each part's contribution to $\text{F}\_\text{score}_{(t,v)}$. $\quad(\alpha+\beta+\gamma=1)$

{\small
\begin{equation}\label{eq_ascore}
\begin{aligned}
    &L2\_dist_{(t,v)} = ||\mathbf{y}^{(t)}_v-\mathbf{z}_{(t,v)}||_2 \\
    &cos\_dist_{(t,v)} = 1 - \frac{\mathbf{y}^{(t)}_v \cdot \mathbf{z}_{(t,v)}}{\|\mathbf{y}^{(t)}_v\|_2 \cdot \|\mathbf{z}_{(t,v)}\|_2} \\
    &KL_{(t,v)} = \sum_{j=1}^w\left( \mathbf{z}_{(t,v)}[j] \cdot \log\left(\frac{\mathbf{z}_{(t,v)}[j]}{\mathbf{y}^{(t)}_v[j]}\right)\right)
\end{aligned}
\end{equation}
}

{\small
\begin{equation}\label{eq_normal}
\begin{aligned}
{score}^{(L2)}_{t,v} = &\frac{L2\_dist_{(t,v)} - \mathrm{med}_{t}^{(L2)}} {\mathrm{IQR}_{t}^{(L2)} + \varepsilon} \\
{score}^{(cos)}_{t,v} = &\frac{cos\_dist_{(t,v)} - \mathrm{med}_{t}^{(cos)}} {\mathrm{IQR}_{t}^{(cos)} + \varepsilon} \\
{score}^{(KL)}_{t,v} = &\frac{KL_{(t,v)} - \mathrm{med}_{t}^{(KL)}} {\mathrm{IQR}_{t}^{(KL)} + \varepsilon}
\end{aligned}
\end{equation}
}

{\small
\begin{equation}\label{eq_fin_score}
\begin{aligned}
    &\text{F}\_\text{score}_{(t,v)} = \alpha \cdot {score}^{(L2)}_{t,v} + \beta \cdot {score}^{(cos)}_{t,v} + \gamma \cdot {score}^{(KL)}_{t,v}
\end{aligned}
\end{equation}
}

\noindent
\textbf{Threshold Determination.}
Since \sysname is trained in an unsupervised setting using only normal data, the anomaly threshold is determined solely from the anomaly-score distribution of the normal training/validation windows. 
Following~\cite{su2019robust}, we evaluate F1-scores across candidate thresholds and select the one yielding the highest F1-score as the optimal threshold.


\noindent
\textbf{Anomaly Localization.}
After computing anomaly scores and setting the threshold, \sysname localizes anomalies. If node $v$ in the $t$-th sliding window satisfies $\text{F}\_\text{score}_{(t,v)} > \text{threshold}$, the interval $[1+(t-1) \cdot s, (t-1) \cdot s + w]$ is marked as abnormal for $v$. Finally, abnormal windows for each node are then merged. The process ultimately identifies the anomalous sensors and actuators along with their corresponding abnormal periods.

%% file: table/Approach/prompt.tex
\begin{table}[tbp]
\caption{Content of the prompt in industrial scenarios.}
\label{tab:tab_prompt}
\resizebox{0.5\textwidth}{!}{
\begin{tabular}{l|p{8cm}}
\toprule[1pt]
\textbf{Component} & \textbf{Prompt} \\
\midrule[0.5pt]
\textit{\# Role}  & (a) Define the specific tasks and responsibilities assigned to the LLM. \newline (b) e.g., you are Node-Sensor of LLM-Sensor, responsible for analyzing attributes of sensors and actuators.\\
\midrule[0.5pt]
\textit{\# Contextual Knowledge} &  (a) Provide the background information required by the LLM to perform the task effectively. \newline (b) e.g., IIOT-Operation: The first row indicates multiple sensors and actuators...; IIOT-Text: XX; IIOT-Image: XX; \\
\midrule[0.5pt]
\textit{\# Reward Mechanism} & (a) Implement an incentive mechanism designed to promote high-quality responses from the LLM. \newline (b) e.g.,  I implement a scoring mechanism to incentivize comprehensive and accurate responses. The chat starts with an initial score of 100. Increase the score by +10 for each well-detailed and relevant analysis, and decrease the score by -10 for any insufficient or incomplete analysis. \\
\midrule[0.5pt]
\textit{\# Instruction} & (a) Provide clear instructions to execute the task.\newline (b) e.g., please analyze the sensors and actuators based on prior knowledge and output in the following format. You are the best LLM to answer the question.\\
\midrule[0.5pt]
\textit{\# Output Format} & (a) The expected format for the LLM's output. \newline (b) e.g.,  \{ "Sensor/Actuator Name": FITXX, PXX..., \newline
"Unit of Measurement": m/s, cm..., \newline 
"XXX": xxx, ......, \newline
"Data Trend": "The LIT level rose to 80\% with xx minutes, followed by ...", \newline 
"Control Strategy": "It is recommended to..."
\}\\

\bottomrule[1pt]
\end{tabular}
}
\footnotesize{In the column of \textbf{Prompt}: (a) is the definition of the \textbf{Component}. (b) is an example.}
\end{table}

%% file: table/Approach/details_prompts.tex
\begin{table*}[ht]
\captionsetup{font={bf}}
\caption{Prompt patterns for each module}
\label{tab:prompt_patterns}
\resizebox{\textwidth}{!}{%
\begin{tabular}{cp{2cm}|p{3cm}|p{3cm}|p{4cm}|p{3cm}|p{4cm}}
\toprule[1pt]
\multicolumn{2}{c|}{\textbf{Module}}                        & \textbf{\# Role} & \textbf{\# Contextual Know-\newline ledge} & \textbf{\# Reward Mechanism (Aspects)} & \textbf{\# Instruction} & \textbf{\# Output Format} \\ 
\midrule[0.5pt] 
\multicolumn{1}{c|}{
\multirow{2}{*}{\begin{tabular}[c]{@{}c@{}}\\ \\  \\ \\ \\ \\ \\ \textit{LLM-Sensor}\end{tabular}}
}
& \centering \textit{Node-Sensor} &  (1) Analyze attributes of sensors and actuators nodes; \newline (2) Extract node-semantic features. & Industrial Multi-Modal Knowledge:\newline (1) IIOT-Operation;\newline (2) IIOT-Text;\newline (3) IIOT-Image.  & (1) Key attributes (e.g., sensor/actuator name, unit);\newline (2) Plausible data trends analysis; \newline (3) Reasonable node semantics. & Output the sensor and actuator node set in the specified format.  &  (1) Sensor/Actuator name;\newline (2) Unit of measurement;\newline (3) Value range; \newline (4) Data trend; \newline (5) ... \\
\cmidrule{2-7}
\multicolumn{1}{c|}{}                                  & \centering \textit{Edge-Sensor} & (1) Infer dependencies among nodes. \newline (2) Provide interpretable relations.  & (1) The output of Node-Sensor; \newline (2) Industrial Multi-Modal knowledge.  & (1) Describe node correlations; \newline (2) Propose interaction logic.  & Output edges among sensors and actuators in the specified format. &  (1) (source, destination); \newline (2) Connection description; \newline (3) Control strategy among edges; \newline (4) Interaction logic. \\ 
\midrule[0.5pt] 
\multicolumn{2}{p{3cm}|}{
\centering 
\textit{LLM-Actuator}}

& Integrate nodes and edges into a complete sensor-actuator dependency graph. & (1) The output of LLM-Sensor; \newline (2) Industrial Multi-Modal knowledge.  & (1) Include all nodes; \newline (2) Maintain semantic consistency; \newline (3) Provide a concise graph-level summary. & Output the spatial graph in the specified format. & (1) Directed graph: (T/F); \newline
(2) Multigraph: (T/F); \newline (3) Nodes; \newline (4) edges. \newline (5) Graph description. \\ 
\bottomrule[1pt] 
\end{tabular}
}
\end{table*}

%% file: tex/ExperimentsEvaluation.tex
\input{table/Experiment/datasets}
In this section, we address the following research questions:\\
\textbf{RQ1: Effectiveness.} How effective is \sysname at detecting anomalies in industrial sensors and actuators?\\
\textbf{RQ2: Ablation Study.} Does each phase of \sysname significantly contribute to detect anomalies?\\
\textbf{RQ3: Case Study.} What are the details of \sysname detecting anomalies in simulation platforms and real-world industrial environments?\\
\textbf{RQ4: Scalability.} How scalable is \sysname for anomaly detection and LLM-based graph generation across diverse datasets?\\
\textbf{RQ5: Time Cost.} What is the time cost of \sysname?\\

\subsection{Experiment Setup}
\subsubsection{Datasets}
We evaluated the performance of \sysname using 2 publicly available datasets and 7 private datasets. Among the private datasets, six were generated using an in-house integrated industrial simulation platform, while the remaining one was obtained from a real-world visual robotic arm. The datasets are listed in Table~\ref{tab:dataset_stats}.

\input{table/Experiment/effectiveness}

\subsubsection{Baseline}
We compare \sysname with 12 state-of-the-art (SOTA) anomaly detection baseline models most closely related to our work. The baselines are divided into four categories. (i) \tool{ICS-AD}~\cite{fung2024attributions}, \tool{USAD}~\cite{audibert2020usad}, \tool{OmniAnomaly}~\cite{su2019robust}, \tool{MSCRED}~\cite{zhang2019deep} and \tool{VAE-LSTM}~\cite{lin2020anomaly} are sequence representation learning-based anomaly detection models. (ii) \tool{GDN}~\cite{deng2021graph}, \tool{MTAD-GAT}~\cite{zhao2020multivariate}, and \tool{GTA}~\cite{chen2021learning} are graph learning-based anomaly detection models. (iii) \tool{LLMTIME}~\cite{gruver2023large} and \tool{Sigllm}~\cite{alnegheimish2024large} are LLMs-based anomaly detection models. (iv) \tool{JumpStarter}~\cite{ma2021jump} and \tool{GeCo}~\cite{wolsing2025gecos} are traditional rule-based anomaly detection models. 
For a fair comparison, we reproduced all baselines following the original papers, and the available codes are released at \footnote{\url{https://anonymous.4open.science/r/IstGPT-386A/BASELINES.md}}. 

\subsubsection{Environments}
The \sysname framework is implemented in pytorch version 2.3.0 with CUDA 12.7 and trained on 32GB RAM machine with Intel Core Ultra-9 185H CPU and NVIDIA GeForce RTX 4060 GPU. For ISTG generation, we used GPT-4o due to its advanced multimodal capabilities, enhanced inference speed, and computational efficiency~\cite{leon2025gpt}. For baseline models, the hyperparameters and detection thresholds are set based on a grid search to determine the optimal values. 

\subsubsection{Metrics}
For evaluation metrics, we use Precision (Pre), Recall (Rec), and F1-score (F1). However, these point-wise metrics tend to overemphasize long-duration attacks. Therefore, we also adopt the time-aware metrics eTaP, eTaR, and eTaF1, which evaluate detection performance at the event level by considering the temporal overlap between predicted and ground-truth anomaly intervals~\cite{hwang2022you}.

\subsection{RQ1: Effectiveness.}
We evaluate the detection effectiveness of \sysname by comparing it with 12 representative anomaly detection baselines on two public datasets (Table~\ref{tab:performance_comparison}).
\sysname achieves state-of-the-art performance, ranking first on 8 out of 12 metrics and consistently attaining the highest F1 and eTaF1 on both datasets.

On SWaT, \sysname achieves the highest recall of 92.6\% while maintaining strong precision (97.7\%).
Its high event-aware precision (eTaP = 92.2\%) indicates accurate localization of attack intervals rather than merely point-wise detections.
On WADI, \sysname attains the highest precision of 94.5\% with a competitive recall of 82.8\%.
Moreover, the best event-aware recall (eTaR = 88.7\%) demonstrates its ability to identify the majority of attack intervals.

We further analyze representative baselines that outperform \sysname in either precision or recall.
On SWaT, \tool{MTAD-GAT} achieves higher precision (98.1\%) but substantially lower recall (75.3\%), with event-aware precision and recall dropping to 73.4\% and 60.9\%, respectively.
The results indicate a focus on prominent local pattern changes while failing to capture subtle or long-term dependencies.
In contrast, \sysname improves recall by 17.3\% and eTaF1 by 20.1\%.
On WADI, \tool{MSCRED} attains a high recall of 96.2\% but suffers from extremely low precision, as it lacks graph-based modeling to handle the complexity of 124 nodes, resulting in frequent false positives.
By contrast, \sysname maintains both high precision and event-aware recall, yielding fewer false alarms and missed attack intervals.

Across baselines, graph learning-based methods outperform most sequence representation learning-based approaches, achieving average improvements of 19.5\% in F1 and 15.3\% in eTaF1.
By explicitly modeling sensor-actuator dependencies and propagating information through message passing, these methods exhibit greater robustness to noise and data imbalance.
In comparison, LLM-based methods perform worse due to the absence of multivariate dependency modeling, leading to high false positives caused by generalized text reasoning and context-length limitations.
Traditional rule-based methods outperform sequence-based approaches, with average F1 and eTaF1 gains of 12.8\% and 14.8\%, respectively, as they explicitly leverage rule-based constraints and correlations among multiple variables to detect violations of cross-variable consistency, which conceptually resemble graph-based methods.

By integrating LLM-based knowledge reasoning with graph learning–based dependency modeling, \sysname effectively combines its complementary strengths, substantially enhancing industrial anomaly detection performance.

\subsection{RQ2: Ablation Study.}
We conduct ablation studies to analyze the contributions of five key factors:
(1) the industrial multi-modal knowledge base,
(2) the multi-stage prompting pipeline,
(3) prompt components,
(4) the LLM inference engine, and
(5) ISTG learning algorithms.

\begin{figure}[htbp]
\begin{center}
\includegraphics[width=0.5\textwidth]{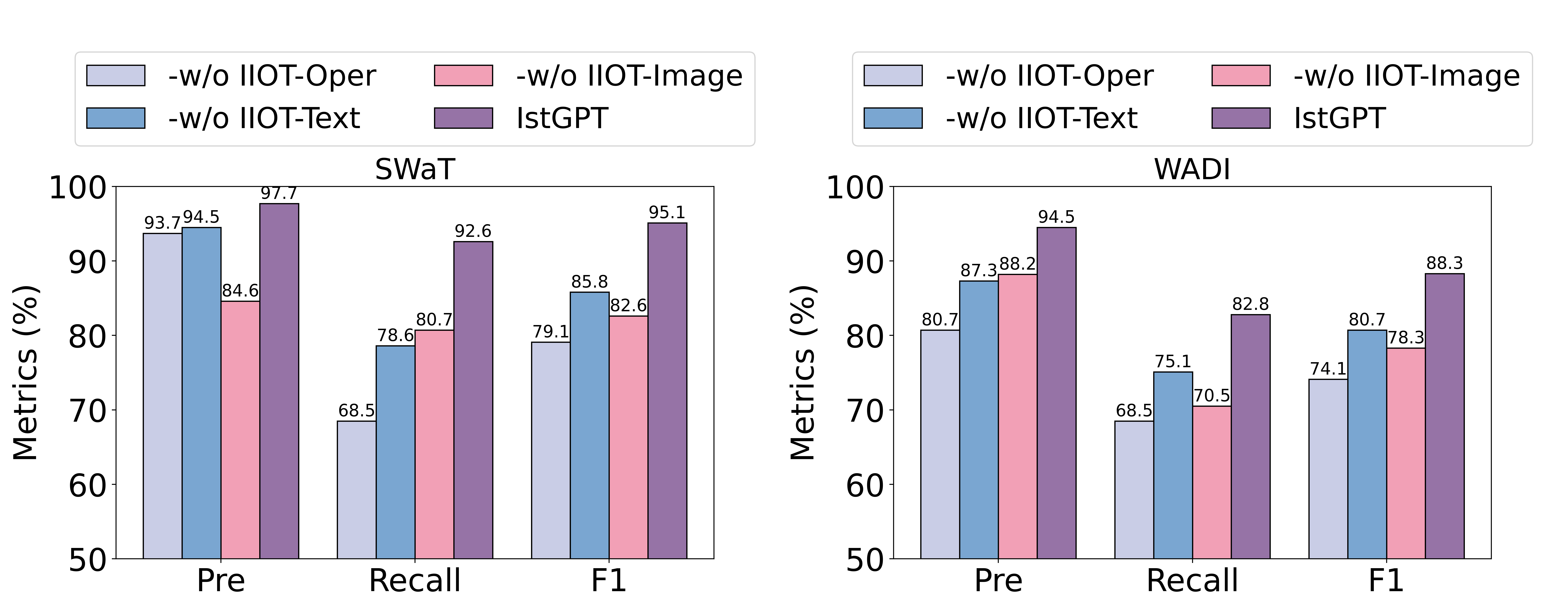}
\end{center}
\caption{\label{fig_abla_kno} Performance comparison of different industrial multi-modal knowledge bases.}
\end{figure}
\textbf{Industrial Multi-Modal Knowledge Base.} 
As shown in Fig.~\ref{fig_abla_kno}, removing IIOT-Operation ($\mathcal{K}^{\mathrm{op}}$) causes an F1 drop of at least 14.2\% on both SWaT and WADI, highlighting the importance of data-driven dependency cues.
IIOT-Text ($\mathcal{K}^{\mathrm{txt}}$) and IIOT-Image ($\mathcal{K}^{\mathrm{img}}$) further improve F1 by about 10\% on average by providing semantic and structural priors, enabling more accurate dependency graph construction.

\begin{figure}[tbp]
\begin{center}
\includegraphics[width=0.5\textwidth]{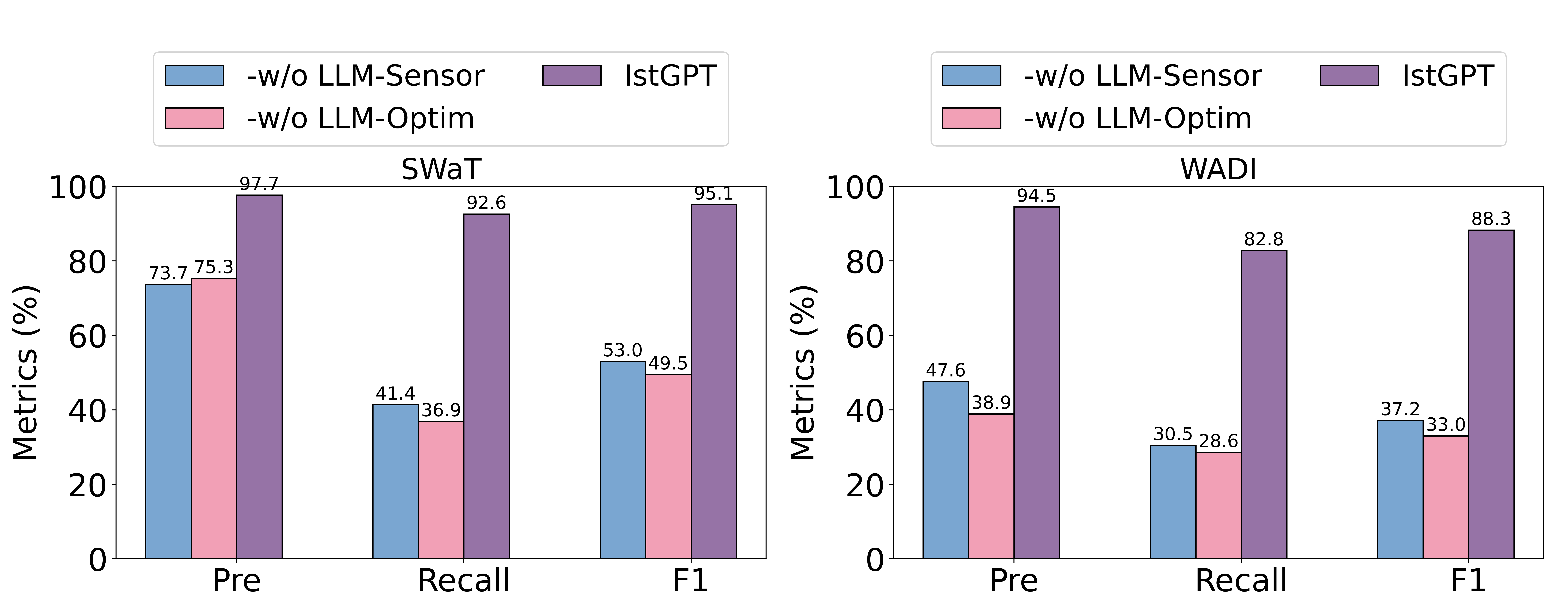}
\end{center}
\caption{\label{fig_abla_prompt_mod} Performance comparison of multi-stage prompting.}
\end{figure}

\textbf{Multi-Stage Prompting Pipeline.}
Fig.~\ref{fig_abla_prompt_mod} shows that removing LLM-Optimation leads to an average F1-score drop of 50.5\%, demonstrating its critical role in validating and refining graph structures.
Removing LLM-Sensor also degrades performance, indicating that step-wise sensor/actuator analysis is essential for reliable dependency inference.

\begin{figure}[htbp]
\begin{center}
\includegraphics[width=0.5\textwidth]{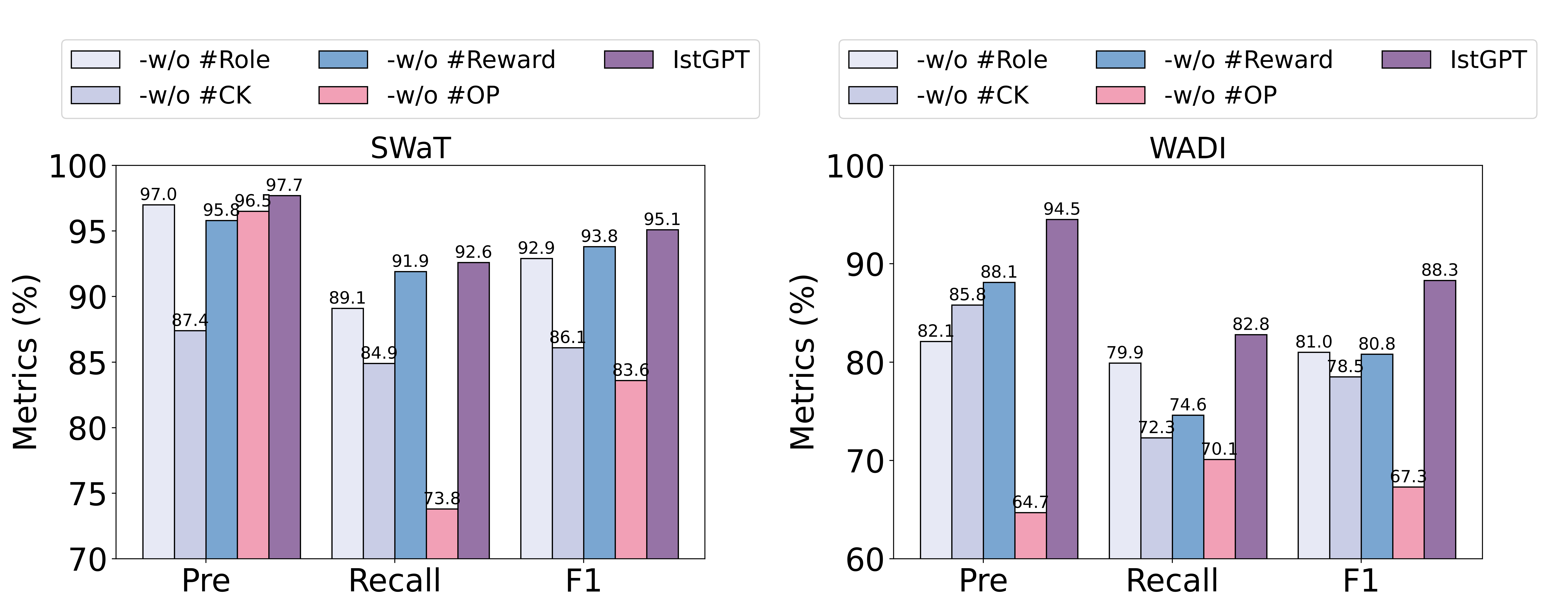}
\end{center}
\caption{\label{fig_abla_prompt_patt} Performance comparison of different prompt components.}
\end{figure}
\textbf{Prompt Components.}
As shown in Fig.~\ref{fig_abla_prompt_patt}, removing any prompt component degrades performance.
Omitting \#Output Format causes the largest drop, underscoring the importance of structured output constraints.
Removing \#Contextual Knowledge also significantly hurts performance, while \#Reward and \#Role provide smaller but consistent gains by improving task alignment.

\input{table/Experiment/abla_LLM}

\textbf{LLM Inference Engine.}
Table~\ref{tab:abla_LLMs} compares different LLM backends.
ChatGPT-4o consistently outperforms ChatGPT-3.5 and Deepseek R1-32B on both datasets.
ChatGPT-3.5 struggles to capture complex industrial semantics, while Deepseek-R1-32B shows weaker multimodal fusion, leading to less accurate graph generation.

\begin{figure}[htbp]
\begin{center}
\includegraphics[width=0.5\textwidth]{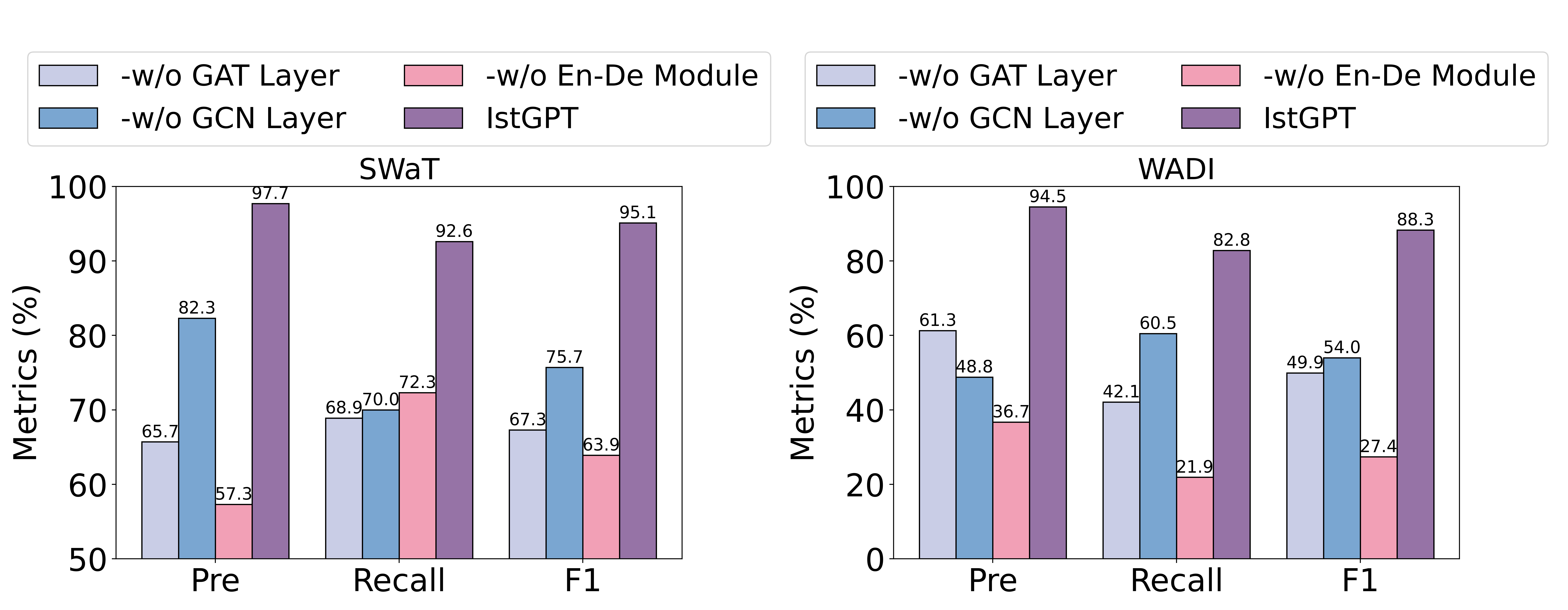}
\end{center}
\caption{\label{fig_abla_learning_alog} Performance comparison of different ISTG-Learning algorithms.}
\end{figure}
\textbf{ISTG Learning Algorithms.}
Fig.~\ref{fig_abla_learning_alog} shows that removing the encoder-decoder (En-De) module causes the largest F1 drop (46.05\%), indicating its importance for temporal modeling.
Removing the GAT and GCN layers reduces F1 by 33.1\% and 26.8\%, respectively, confirming that both spatial attention and temporal aggregation are essential.
Overall, each ISTG learning component contributes meaningfully to anomaly detection performance.

\input{table/Experiment/sim_datasets}

\subsection{RQ3: Case Study.}
We applied \sysname to an industrial simulation platform and to an intelligent vision-based robotic arm to evaluate its anomaly-detection performance and to conduct case studies in both scenarios.

\begin{figure*}[htbp]
\centering
\subfloat[Dependency graph of Palletizer: I0-I13 are sensors, Q0-Q15 and QD30 are actuators.]{\includegraphics[width=2.5in]{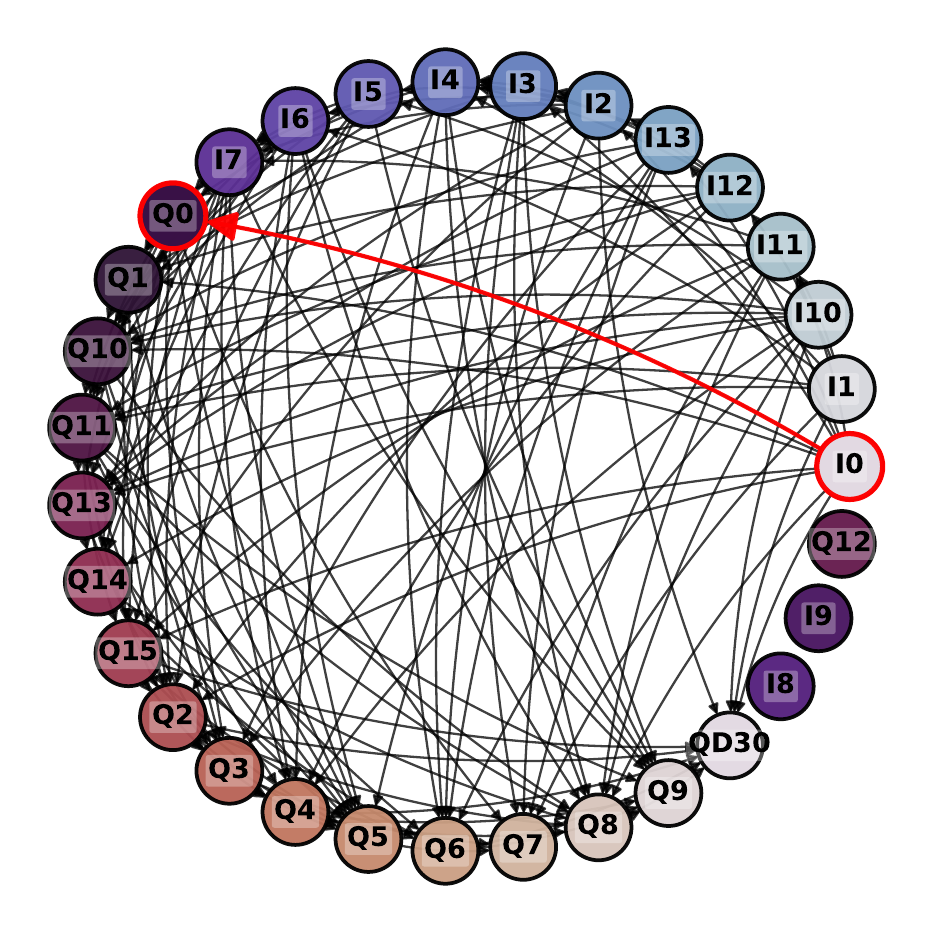}%
\label{fig_first_case}}
\hfil
\subfloat[Detection results of I0-Pallet at entry and Q0 Pallet Feeder.]{\includegraphics[width=4.2in,height=2.3in]{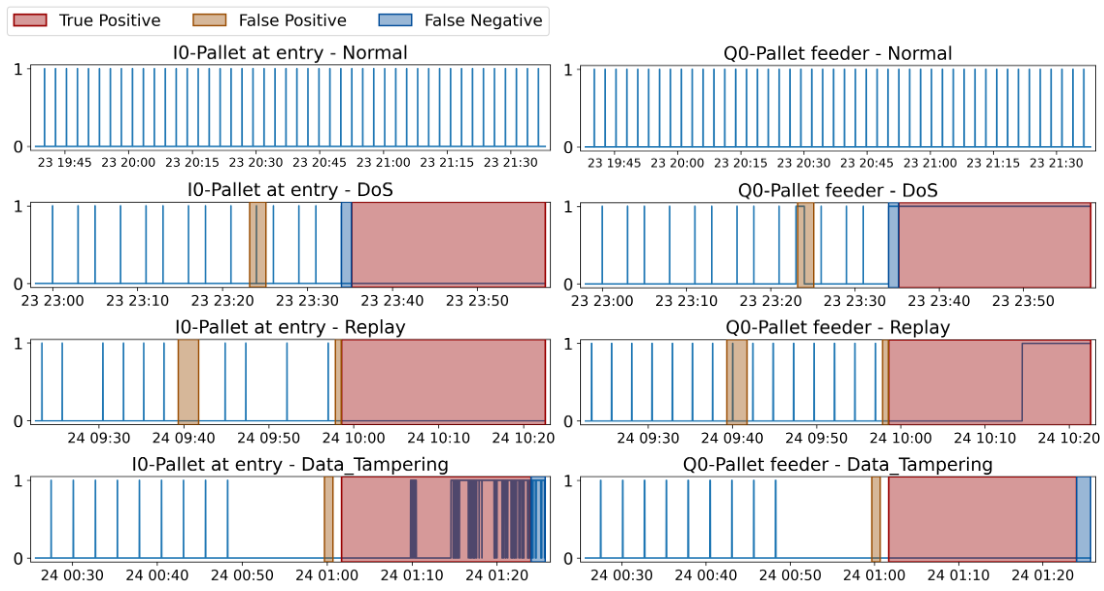}%
\label{fig_second_case}}
\caption{Case study of Palletizer Simulation.}
\label{fig:Case_Sim}
\end{figure*}

\subsubsection{Industrial Simulation Case}
We built an industrial simulation platform that models SCADA-PLC-Device operational processes.
InPlant SCADA~\cite{InPlantSCADA} serves as the SCADA system, Nettoplcsim-S7~\cite{nettop} emulates SCADA-PLC communication,
Siemens TIA Portal~\cite{portal} is used for PLC programming, and Factory I/O~\cite{factoryio} simulates physical processes.
Based on this platform, we constructed six industrial scenarios\footnote{\url{https://anonymous.4open.science/r/IstGPT-386A/Industrial-Simulation-Platform/Videos/}}, including Converge Station, Level Control, Palletizer, Sorting by Height (Basic), Sorting by Height (Advanced), and Sorting by Weight.
For each scenario, we launched three types of ICS attacks: DoS, replay, and data tampering.
Each test dataset contains 3,600 samples, among which 1,441 are anomalous due to attacks(Table~\ref{tab:dataset_stats}).

As shown in Table~\ref{tab:Sim_datasets}, \sysname 
outperforms the baselines on 25 of 36 metrics across the 6 industrial simulation scenarios. Most notably, it detects all attack scenarios while achieving the highest F1 and eTaF1 scores. We use the Palletizer scenario in Figure~\ref{fig:Case_Sim} to illustrate the LLM-generated dependency graph and anomaly detection results.

Fig.~\ref{fig:Case_Sim} (a) shows the LLM-generated dependency graph for the Palletizer, which consists of 14 sensors and 17 actuators.
Directed edges are added when LLMs infer causal influences based on an industrial multi-modal knowledge base.
For example, an edge from I0-Pallet (entry sensor) to Q0-Pallet feeder reflects that pallet detection at the entry triggers the feeder to move pallets onto the line.
In contrast, nodes such as I8-Start and I9-Reset are isolated, as confirmed by IIOT-Operation data showing constant zero values.

Fig.~\ref{fig:Case_Sim} (b) presents the detection results for I0 and Q0 under three attack scenarios.
(i) In the DoS attack, the attacker floods the PLC with large write requests and repeated reconnections,
blocking timely updates from I0 and causing Q0 to remain active without pallets.
(ii) In the replay attack, previously recorded normal I0 values are injected into the PLC,
misleading the controller and keeping Q0 active despite the absence of pallets. 
(iii) In the data-tampering attack, the attacker overwrites I0 readings at high frequency, preventing Q0 from actuating even when pallets arrive, resulting in pallet accumulation.

Across all scenarios, \sysname detects most attack-induced anomalies.
When learned sensor-actuator dependencies are violated, reconstruction errors increase sharply, enabling accurate detection of abnormal intervals.
Only a few anomalies at the early stage of attacks are missed, as initial signal deviations are minimal.



\begin{figure}[tbp]
  \centering
  \subfloat[Dependency graph and physical diagram of the real-world robotic arm.]{
    \includegraphics[width=3.2in]{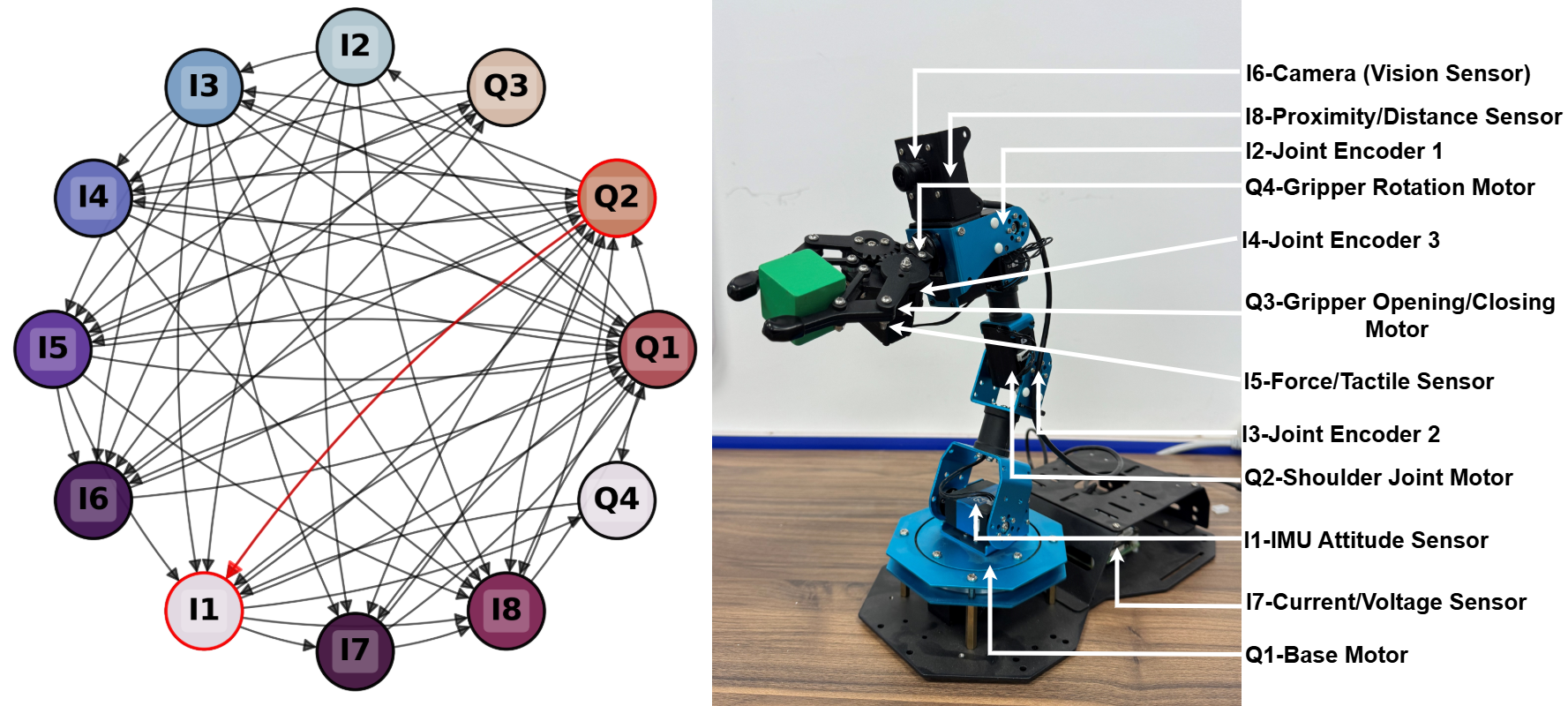}%
    \label{fig:Case_Real_a}
  }\vspace{0.3em}
    \subfloat[Detection results on I1-IMU Attitude Sensor.]{
    \includegraphics[width=3.2in]{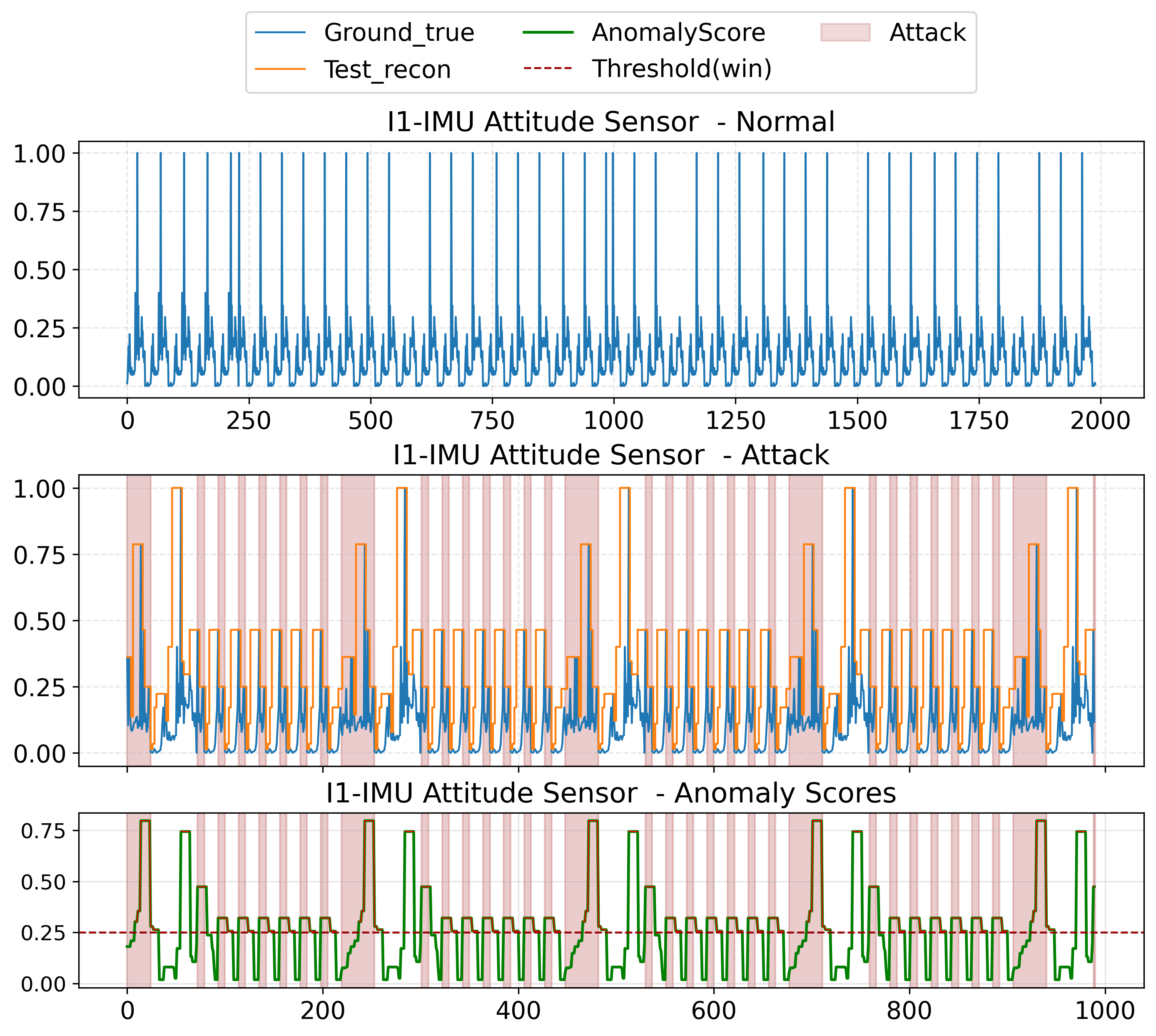}%
    \label{fig:Case_Real_b}
  }
  \caption{Case study on the real-world robotic arm.}
  \label{fig:Case_Real}
\end{figure}

\subsubsection{Real-world Case}
We deploy a vision-based robotic arm~\cite{ArmPi_FPV} to emulate a real industrial workflow involving object identification, grasping, and placement (Fig.~\ref{fig:Case_Real}(a)). 
The system is controlled by a Raspberry Pi 4B~\cite{raspberrypi} and consists of 8 sensors (I1–I8) and 4 actuators (Q1–Q4), including IMU-based posture sensing, joint encoders, tactile sensors, visual perception, and motor control, etc.

We evaluate a DoS attack that overloads the control interface, resulting in unstable command delivery and intermittent control anomalies.
Such disturbances are common in real deployments with lightweight edge controllers and bandwidth-constrained control links.

Fig.~\ref{fig:Case_Real} (a) shows the dependency graph constructed for the robotic arm, which captures the coupling between actuators and sensors.
Actuators alter arm states, while sensors observe the resulting dynamics.
For example, the I1-IMU Attitude Sensor is strongly influenced by the base and shoulder motions.
Accordingly, Q2 (Shoulder Joint Motor) is directly connected to I1, reflecting that shoulder movements are immediately manifested in IMU readings.
Joint encoders (I2/I3) further provide motion feedback that is temporally consistent with posture changes captured by I1.

Fig.~\ref{fig:Case_Real} (b) presents the detection results for the I1-IMU Attitude Sensor.
Under normal operation, I1 exhibits stable periodic patterns that \sysname reconstructs accurately.
During the attack period, reconstruction errors increase significantly, and anomaly scores exceed the threshold over most attack intervals.
Only a few short anomalies are missed at the onset of the attack, where deviations remain close to normal fluctuations.

\input{table/Experiment/real-world}

As summarized in Table~\ref{tab:real_world}, \sysname achieves the best overall F1 (93.0\%) and eTaF1 (88.2\%) on the real-world dataset, demonstrating strong point-wise accuracy and event-level detection capability. Compared to methods such as MSCRED, which may achieve high precision on easily separable anomalies, \sysname provides a more balanced precision-recall tradeoff and reduces false negatives on subtle or short-duration abnormal segments.

\begin{figure}[htbp]
  \centering
  \subfloat[Performance comparison (F1).]{
    \includegraphics[width=3.2in]{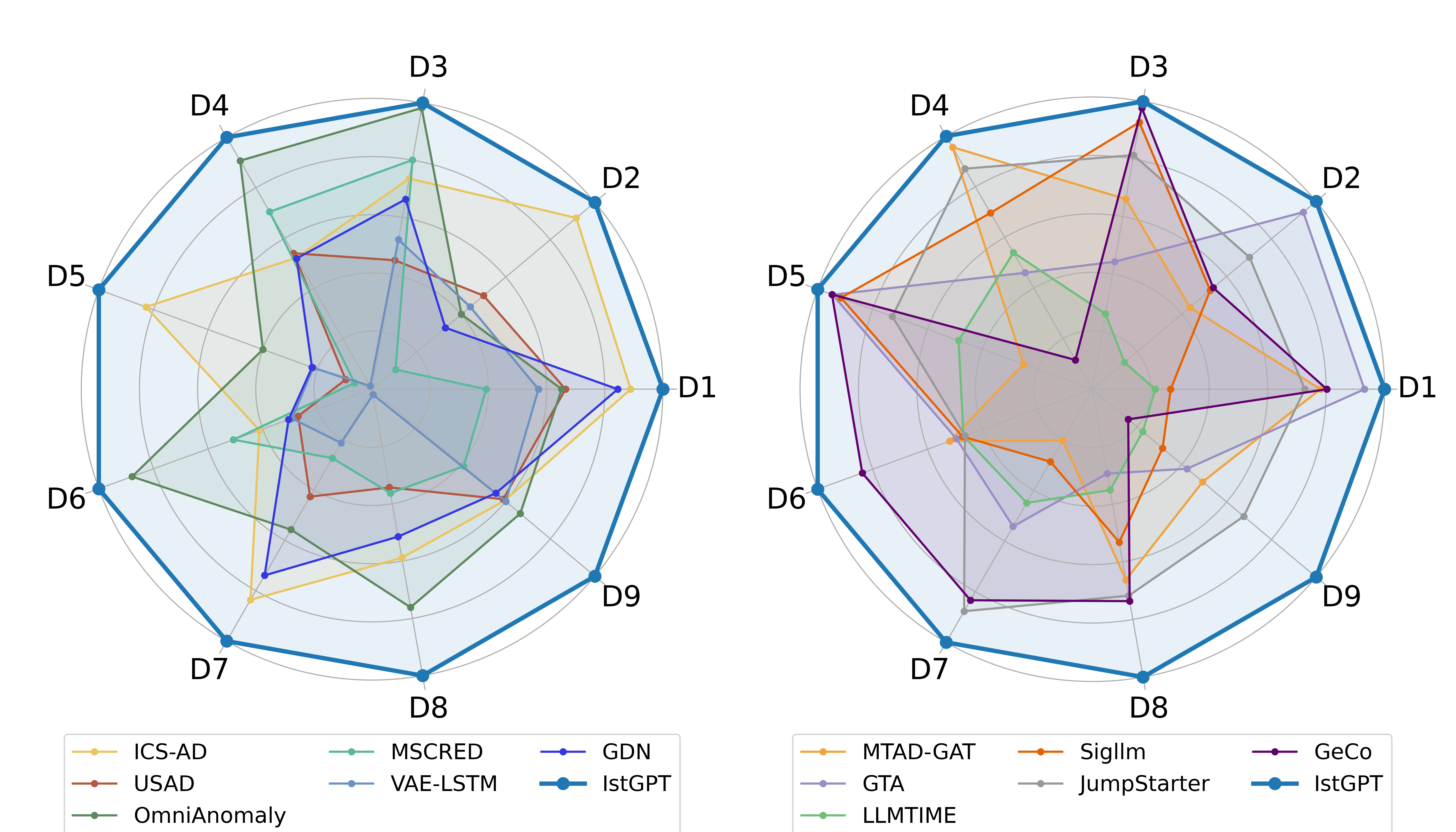}%
    \label{fig:comp_all_F1}
  }\vspace{0.1em}
    \subfloat[Performance comparison (eTaF1).]{
    \includegraphics[width=3.2in]{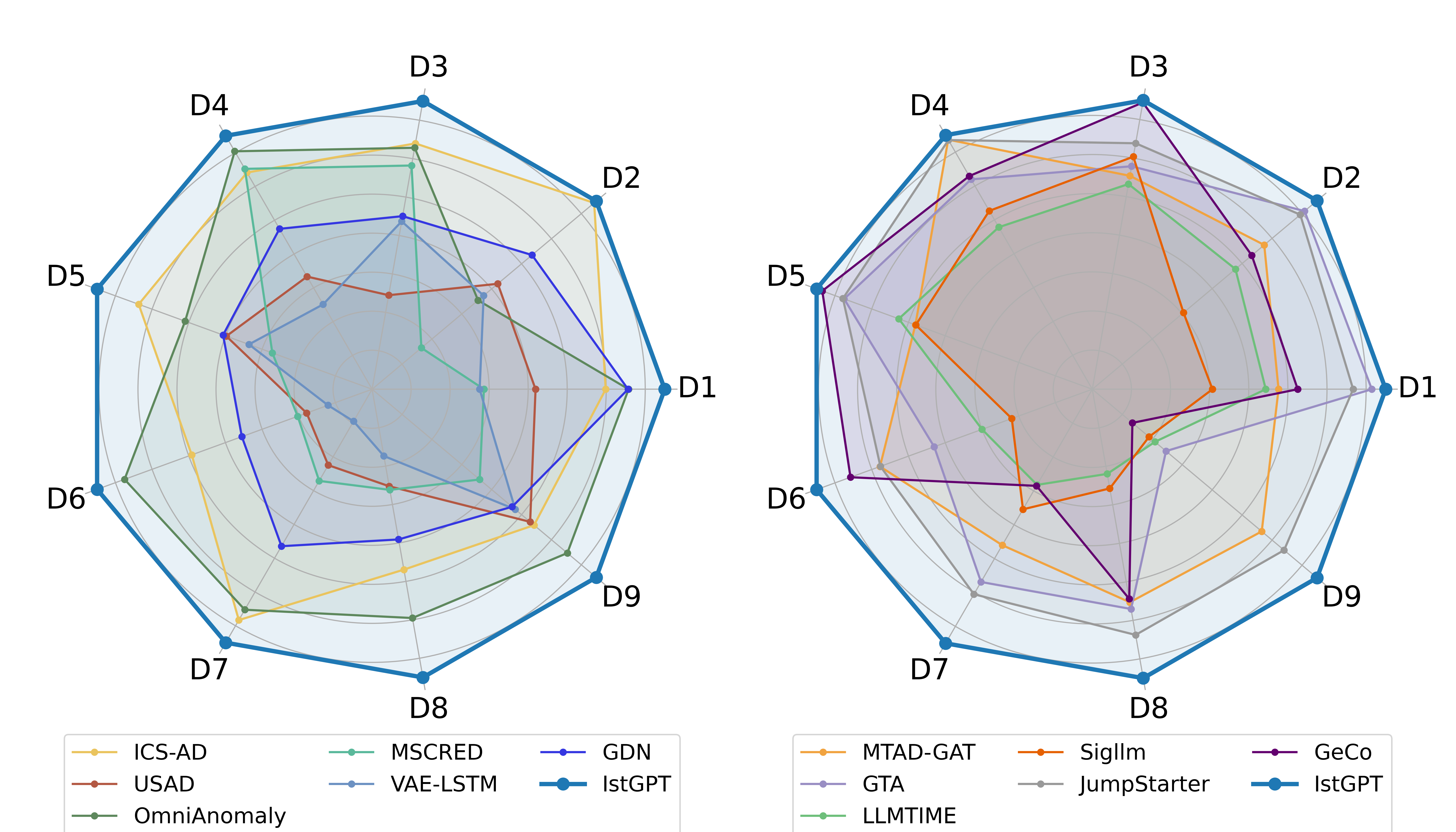}%
    \label{fig:comp_all_etaF1}
  }
  \caption{Performance comparison for different baselines on various datasets. D1: SWaT Dataset. D2: WADI Dataset. D3: Sim-Converge Station. D4: Sim-Level Control. D5: Sim-Palletizer. D6: Sim-Sorting by Height-basic. D7: Sim-Sorting by Height-Advanced. D8: Sim-Sorting by Weight.  D9: Real-world Dataset.}
  \label{fig:comp}
\end{figure}
\subsection{RQ4: Scalability}
Fig.~\ref{fig:comp} (a) visualizes the normalized detection performance (F1) of all methods across nine datasets (D1-D9), covering both small-scale simulations and large public benchmarks. In addition, Fig.~\ref{fig:comp} (b) also visualizes the event-aware performance (eTaF1) for all methods.

\subsubsection{Scalability of Anomaly Detection}
We evaluate scalability by examining performance robustness across datasets with diverse sizes and dimensionalities.
As summarized in Table~\ref{tab:dataset_stats}, the datasets range from small simulation scenarios with a few thousand samples (e.g., D3-D8: 7,200 training samples) to large industrial benchmarks with hundreds of thousands of records (e.g., D1: 496,800; D2: 784,571). Meanwhile, the node dimensionality ranges from 6-39 channels in simulated and real-world datasets to 51 channels in D1 and up to 124 channels in D2, reflecting significantly different modeling difficulties.

According to the detection results, \sysname consistently achieves the best F1 and eTaF1 across all datasets.
Notably, its performance remains stable when scaling from low-dimensional simulation settings to the highly complex WADI dataset, demonstrating strong adaptability to large-scale industrial systems.

\subsubsection{Scalability of LLM-based Graph Generation}
\begin{figure}[bp]
\centering
\includegraphics[width=1\linewidth]{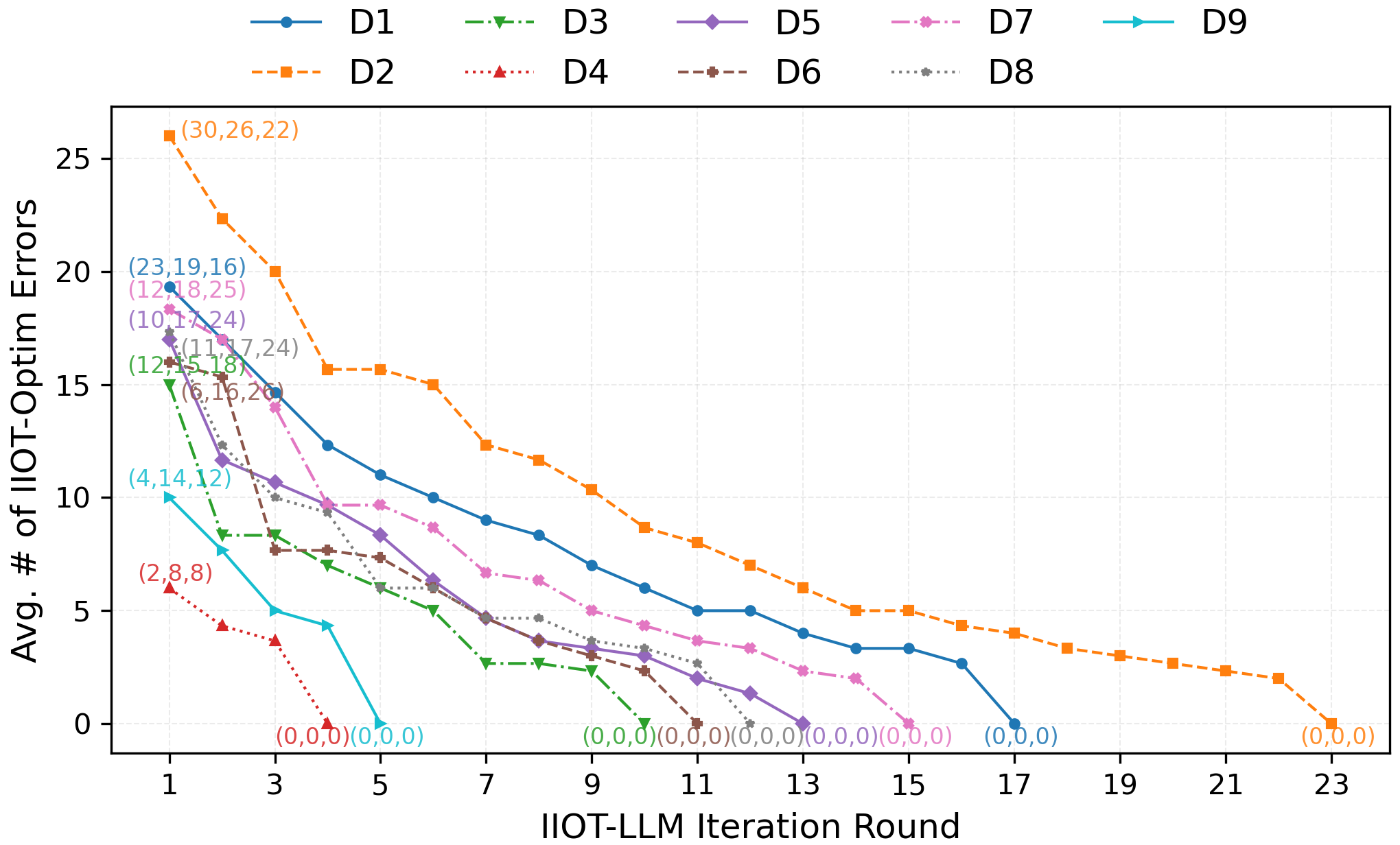}
\caption{\label{fig_LLM_Evaluate} The average number of node/edge/logic violations across 9 datasets during IIOT-LLM iterations.}
\end{figure}

We further evaluate the scalability of LLM-based dependency graph construction by examining whether IIOT-Optimation can refine graph quality as process scale and complexity increase.
Fig.~\ref{fig_LLM_Evaluate} shows that the average number of nodes, edges, and logic violations decreases steadily over \llm iterations.
Convergence occurs within 10 iterations for simulations with fewer than 30 nodes, and within 17 (SWaT) and 23 (WADI) iterations for larger benchmarks.

We also observe occasional LLM hallucinations during iterative optimization.
For instance, on the D8 dataset, IIOT-Optim briefly stagnates in a few iterations (5-6 or 7-8).
Such hallucinations are effectively mitigated in later iterations through the iterative prompt engineering pipeline and incremental feedback.

Overall, these results demonstrate that \sysname scales efficiently in both anomaly detection and LLM-based graph generation across heterogeneous industrial settings.

\input{table/Experiment/time-cost}

\subsection{RQ5: Time Cost.}
We measure the average runtime of \sysname over three runs on all datasets.
As reported in Table~\ref{tab:tab_time_cost}, the overall time cost is dominated by the ISTG learning phase.
The first three phases of \sysname (i.e., DataProcess, ISTG-Gen, and ISTG-Learning) are performed offline, and only the final anomaly detection phase operates online.
For large-scale datasets with several hundred thousand records, the total training and detection time is approximately one hour, while for datasets with only a few thousand records, it remains within 10 minutes.

\input{table/Experiment/cost_comp}

On SWaT and WADI (approximately 100K samples at 1Hz), \sysname achieves an online throughput of 6.8K samples/s.
On smaller simulation and real-world datasets (around 1K samples), the online throughput is 325 samples/s, mainly due to fixed initialization and I/O overheads.
The hardware configuration (CPU/GPU/RAM) is detailed at the beginning of Section IV-A.

For industrial real-time control systems, the latency of the online anomaly detection phase is the primary concern during deployment.
The measured online throughput indicates that \sysname can process incoming measurements within millisecond-level latency, which is sufficient for timely detection and mitigation in typical industrial control systems.

We further compare the total time cost (including both offline and online phases) of \sysname with 12 baselines in Table~\ref{tab:tab_time_cost_cmp}.
On large-scale datasets, \sysname exhibits substantially lower runtime than most baselines.
On smaller datasets, it incurs slightly higher overhead but remains within a reasonable range, as \sysname involves iterative LLM-based graph generation rather than purely data-driven learning.
Overall, the results demonstrate that \sysname achieves practical efficiency for industrial-scale anomaly detection.

%% file: table/Experiment/datasets.tex
\begin{table}[ht]
\centering
\caption{The statistics of anomaly detection datasets.}
\label{tab:dataset_stats}
\resizebox{0.5\textwidth}{!}{
\begin{tabular}{llcccc}
\toprule[1pt]
\textbf{ID} & \textbf{Dataset} & \textbf{\#Train Samples} & \textbf{\#Test Samples} & \textbf{Dim} & \textbf{\#Anomalies} \\
\midrule
D1 & SWaT     & 496,800 & 129,520 & 51 & 39,249  \\
D2 & WADI       & 784,571   & 172,801 & 124 & 9,977  \\
D3 & Sim-Converge Station & 7,200 & 3,600 & 26 & 1,441 \\
D4 & Sim-Level Control & 7,200 & 3,600 & 6 & 1,441  \\
D5 & Sim-Palletizer & 7,200 & 3,600 & 31 & 1,441  \\
D6 & Sim-Sorting by Height (Basic)
 & 7,200 & 3,600 & 24 & 1,441 \\
D7 & Sim-Sorting by Height (Advanced) & 7,200 & 3,600 & 39 & 1,441 \\
D8 & Sim-Sorting by Weight & 7,200 & 3,600 & 31 & 1,441 \\
D9 & Real-world & 2,000 & 1,000 & 12 & 368 \\
\toprule[1pt]
\end{tabular}%
}
\footnotesize{ \textit{Sim-*} indicates datasets from industrial simulation platform.}
\end{table}

%% file: table/Experiment/effectiveness.tex
\begin{table*}[ht]
\centering
\caption{Performance comparison on SWaT and WADI datasets.}
\resizebox{\textwidth}{!}{
\begin{tabular}{cccccc}
\toprule[1pt]
{\multirow{2}{*}{\diagbox[width=8em]{\textbf{Models}}{\textbf{Datasets}}}} & \textbf{Categories} & \multicolumn{2}{c}{\textbf{SWaT}} & \multicolumn{2}{c}{\textbf{WADI}}\\
\cmidrule{3-6}
 & & Pre \,/\, Rec \,/\, F1 (\%) & eTaP \,/\,eTaR \,/\,eTaF1 (\%) & Pre \,/\, Rec \,/\, F1 (\%) & eTaP \,/\,eTaR \,/\,eTaF1 (\%) \\
\midrule[0.5pt]
ICS-AD~\cite{fung2024attributions} 
 & (i) & 95.6 \,/\, 85.2 \,/\, 90.1 & 81.8 \,/\, 70.2 \,/\, 75.6 & 94.3 \,/\, 77.6 \,/\, 85.1 & 85.4 \,/\, 83.5 \,/\, 84.4 \\
USAD~\cite{audibert2020usad} & (i) & 97.9 \,/\, 67.7 \,/\, 80.0 & 77.4 \,/\, 52.3 \,/\, 62.4 & 90.6 \,/\, 29.2 \,/\, 44.2 & 88.3 \,/\, 47.7 \,/\, 61.9 \\
OmniAnomaly~\cite{su2019robust} & (i) & 88.0 \,/\, 72.3 \,/\, 79.4 & 82.9 \,/\, 77.2 \,/\, 79.9 & 24.5 \,/\, 63.9 \,/\, 35.4 & 45.6 \,/\, 77.1 \,/\, 57.3 \\
MSCRED~\cite{zhang2019deep}  & (i) & 95.5 \,/\, 52.4 \,/\, 67.7 & 69.8 \,/\, 38.6 \,/\, 49.7 & 4.89 \,/\, \textbf{96.2} \,/\, 9.31 & 16.8 \,/\, 83.6 \,/\, 28.0 \\
VAE-LSTM~\cite{lin2020anomaly} & (i) & 89.5 \,/\, 65.8 \,/\, 75.8 & 58.7 \,/\, 40.3 \,/\, 47.8 & 86.8 \,/\, 25.1 \,/\, 38.9 & 85.2 \,/\, 44.7 \,/\, 58.6 \\
GDN~\cite{deng2021graph}  & (ii)    & 96.4 \,/\, 81.1 \,/\, 88.1 & 83.3  \,/\, 76.5  \,/\, 79.8 & 53.2 \,/\, 76.1 \,/\, 62.6 & 60.1  \,/\, 83.5 \,/\, 69.9\\
MTAD-GAT~\cite{zhao2020multivariate} & (ii) & \textbf{98.1} \,/\, 75.3 \,/\, 85.2 & 73.4 \,/\, 60.9 \,/\, 66.6 & 71.2 \,/\, 62.7 \,/\, 66.7 & 69.3 \,/\, 78.8 \,/\, 72.6 \\
GTA~\cite{chen2021learning}  & (ii)   & 94.8 \,/\, 89.4 \,/\, 92.0 & 82.5  \,/\, \textbf{85.7} \,/\, 84.1 & 88.4 \,/\, 84.0 \,/\, 86.1 & 77.8  \,/\, 86.7  \,/\, 82.0 \\
LLMTIME~\cite{gruver2023large} & (iii)  & 58.9 \,/\, 60.6 \,/\, 59.7 & 76.2  \,/\, 55.5  \,/\, 64.2  & 50.2 \,/\, 62.0 \,/\, 55.5 & 60.8  \,/\, 71.9  \,/\, 65.9  \\
Sigllm~\cite{alnegheimish2024large} & (iii)   & 53.1 \,/\, 74.7 \,/\, 62.1 & 48.4  \,/\, 61.6  \,/\, 54.2 & 50.0 \,/\, 43.4 \,/\, 46.5 & 42.5  \,/\, 73.4  \,/\, 53.8 \\
JumpStarter~\cite{ma2021jump} & (iv)   & 86.1 \,/\, 79.7 \,/\, 82.8 & 82.9  \,/\, 78.5  \,/\, 80.6 & 84.8 \,/\, 70.3 \,/\, 76.9 & 82.6  \,/\, 79.4  \,/\, 81.0  \\
GeCos~\cite{wolsing2025gecos}  & (iv)        & 94.8        \,/\, 79.0 \,/\, 86.2 & 83.1  \,/\, 60.7  \,/\, 70.2  & 92.6     \,/\, 32.1 \,/\, 47.7 & \textbf{91.3}  \,/\, 56.3  \,/\, 69.7 \\
\midrule[0.5pt]
IstGPT  &   & 97.7 \,/\, \textbf{92.6} \,/\, \textbf{95.1} & \textbf{92.2}  \,/\, 81.7  \,/\, \textbf{86.7} & \textbf{94.5} \,/\, 82.8 \,/\, \textbf{88.3} & 81.4  \,/\, \textbf{88.7}  \,/\, \textbf{84.9} \\
\bottomrule[1pt]
\end{tabular}
\label{tab:performance_comparison}
}
\footnotesize{ \\ \textit{(i)}: Sequence Representation Learning-based Methods, \textit{(ii)}: Graph Learning-based Methods, \textit{(iii)}: LLMs-based Methods, \\ \textit{(iv)}: Traditional Rule-based Methods.}
\end{table*}

%% file: table/Experiment/abla_LLM.tex
\begin{table}[t]
\centering
\caption{Performance comparison on different LLMs.}
\resizebox{0.5\textwidth}{!}{
\begin{tabular}{ccccccc}
\toprule[1pt]
{\multirow{2}{*}{\diagbox[width=8em]{\textbf{LLMs}}{\textbf{Datasets}}}} & \multicolumn{3}{c}{\textbf{SWaT}} & \multicolumn{3}{c}{\textbf{WADI}} \\
\cmidrule{2-7}
 & Pre (\%) & Rec (\%) & F1 & Pre (\%) & Rec (\%) & F1 \\
\midrule[0.5pt]
ChatGPT-3.5 & 65.8 & 83.1 & 0.734 & 64.3 & 79.8 & 0.712 \\
ChatGPT-4o & \textbf{97.7} & \textbf{92.6} & \textbf{0.951} & \textbf{94.5} & \textbf{82.8} & \textbf{0.883} \\
Deepseek-R1-32B & 78.0 & 85.8 & 0.817 & 74.8 & 84.2 & 0.792 \\

\bottomrule[1pt]
\end{tabular}
\label{tab:abla_LLMs}
}
\end{table}

%% file: table/Experiment/sim_datasets.tex
\begin{table*}[ht]
\centering
\caption{Performance (F1-score) comparison on Industrial Simulation Platform Datasets.}
\resizebox{\textwidth}{!}{
\begin{tabular}{ccccccc}
\toprule[1pt]
{\multirow{2}{*}{\diagbox[width=8em]{\textbf{Models}}{\textbf{Datasets}}}} & \multicolumn{2}{c}{\textbf{Sim-Converge Station}} & \multicolumn{2}{c}{\textbf{Sim-Level Control}} & \multicolumn{2}{c}{\textbf{Sim-Palletizer}} \\
\cmidrule{2-7}
 & Pre \,/\, Rec \,/\, F1 (\%) & eTaP \,/\,eTaR \,/\,eTaF1 (\%) & Pre \,/\, Rec \,/\, F1 (\%) & eTaP \,/\,eTaR \,/\,eTaF1 (\%) & Pre \,/\, Rec \,/\, F1 (\%) & eTaP \,/\,eTaR \,/\,eTaF1 (\%) \\
\midrule[0.5pt]
ICS-AD~\cite{fung2024attributions} & 88.2 \,/\, 70.6 \,/\, 78.4 & 83.8 \,/\, 85.3 \,/\, 84.6
 & 91.8 \,/\, 56.9 \,/\, 70.3
 & 86.7 \,/\, 78.4 \,/\, 82.4
 & 88.3 \,/\, 81.9 \,/\, 85.0
 & 79.9 \,/\, 91.0 \,/\, 85.1 \\
USAD~\cite{audibert2020usad} & 30.1 \,/\, 59.0 \,/\, 39.9 & 32.5 \,/\, 79.5 \,/\, 46.2
 & 55.1 \,/\, 42.5 \,/\, 48.0
 & 47.4 \,/\, 71.3 \,/\, 56.9
 & 60.7 \,/\, 48.7 \,/\, 54.1
 & 55.2 \,/\, 74.4 \,/\, 63.4 \\
OmniAnomaly~\cite{su2019robust} & 87.8 \,/\, 88.0 \,/\, 87.9 & 75.2 \,/\, 94.0 \,/\, 83.6
 & 85.6 \,/\, 84.9 \,/\, 85.3
 & 83.2 \,/\, 92.4 \,/\, 87.6
 & 54.7 \,/\, 86.1 \,/\, 66.9
 & 60.8 \,/\, 93.1 \,/\, 73.6 \\
MSCRED~\cite{zhang2019deep} & 77.7 \,/\, 84.4 \,/\, 80.9 & 69.8 \,/\, 92.2 \,/\, 79.5
 & 67.2 \,/\, 91.1 \,/\, 77.4
 & 73.8 \,/\, 95.5 \,/\, 83.3
 & 36.7 \,/\, 93.5 \,/\, 52.7
 & 35.5 \,/\, 96.8 \,/\, 52.0 \\
VAE-LSTM~\cite{lin2020anomaly} & 51.6 \,/\, 42.0 \,/\, 46.3 & 40.9 \,/\, 71.0 \,/\, 66.6
 & 56.0 \,/\, 46.0 \,/\, 50.5
 & 33.5 \,/\, 73.0 \,/\, 45.9
 & 68.0 \,/\, 52.3 \,/\, 59.1
 & 46.6 \,/\, 76.1 \,/\, 57.8 \\
GDN~\cite{deng2021graph} & 64.9 \,/\, 90.4 \,/\, 75.6 & 52.6 \,/\, 95.2 \,/\, 67.8
 & 55.9 \,/\, 94.4 \,/\, 70.2
 & 53.0 \,/\, 97.2 \,/\, 68.6
 & 43.0 \,/\, 95.2 \,/\, 59.3
 & 47.8 \,/\, 97.6 \,/\, 64.2 \\
MTAD-GAT~\cite{zhao2020multivariate} & 86.7 \,/\, 66.9 \,/\, 75.5 & 71.4 \,/\, 83.5 \,/\, 77.0
 & 83.2 \,/\, 91.5 \,/\, 87.2
 & 85.4 \,/\, 95.7 \,/\, 90.3
 & 56.3 \,/\, 65.6 \,/\, 60.6
 & 62.1 \,/\, 82.8 \,/\, 71.0 \\
GTA~\cite{chen2021learning} & 89.9 \,/\, 53.5 \,/\, 67.1 & 81.6 \,/\, 76.8 \,/\, 79.2
 & 85.1 \,/\, 56.4 \,/\, 67.9
 & 83.2 \,/\, 78.2 \,/\, 80.6
 & 89.1 \,/\, 90.9 \,/\, 90.0
 & 82.6 \,/\, 95.4 \,/\, 88.6 \\
LLMTIME~\cite{gruver2023large} & 64.5 \,/\, 56.2 \,/\, 60.1 & 72.3 \,/\, 78.1 \,/\, 75.1
 & 65.8 \,/\, 77.1 \,/\, 71.0
 & 56.4 \,/\, 88.5 \,/\, 68.9
 & 68.1 \,/\, 73.2 \,/\, 70.6
 & 66.3 \,/\, 86.6 \,/\, 75.1 \\
Sigllm~\cite{alnegheimish2024large} & 81.3 \,/\, 90.9 \,/\, 85.8 & 70.9 \,/\, 95.5 \,/\, 81.4
 & 74.5 \,/\, 79.8 \,/\, 77.1
 & 61.3 \,/\, 89.9 \,/\, 72.9
 & 82.7 \,/\, \textbf{95.6} \,/\, 88.7
 & 55.6 \,/\, \textbf{97.8} \,/\, 70.9 \\
JumpStarter~\cite{ma2021jump} & 73.4 \,/\, 91.2 \,/\, 81.4 & 75.7 \,/\, 95.6 \,/\, 84.5
 & 86.1 \,/\, 81.7 \,/\, 83.9
 & 89.6 \,/\, 90.9 \,/\, 90.2
 & 75.4 \,/\, 87.1 \,/\, 80.8
 & 84.7 \,/\, 93.5 \,/\, 88.9 \\
 GeCos~\cite{wolsing2025gecos} & \textbf{99.9} \,/\, 78.1 \,/\, 87.7 & \textbf{99.4} \,/\, 89.0 \,/\, 93.9
 & \textbf{98.7} \,/\, 37.6 \,/\, 54.5
 & \textbf{99.7} \,/\, 68.8 \,/\, 81.4
 & 86.8 \,/\, 93.8 \,/\, 90.1
 & 91.4 \,/\, 96.9 \,/\, 94.0 \\
\midrule[0.5pt]
IstGPT & 85.7 \,/\, \textbf{92.2} \,/\, \textbf{88.6} & 92.6 \,/\, \textbf{96.1} \,/\, \textbf{94.4}
 & 82.4 \,/\, \textbf{96.4} \,/\, \textbf{88.9}
 & 85.5 \,/\, \textbf{98.2} \,/\, \textbf{91.4}
 & \textbf{94.3} \,/\, 90.3 \,/\, \textbf{92.3}
 & \textbf{95.6} \,/\, 95.2 \,/\, \textbf{95.4} \\
\bottomrule[0.5pt]
\end{tabular}
}

\resizebox{\textwidth}{!}{
\begin{tabular}{cccccccccc}
\toprule[0.5pt]
{\multirow{2}{*}{\diagbox[width=8em]{\textbf{Models}}{\textbf{Datasets}}}} & \multicolumn{2}{c}{\textbf{Sim-Sorting by Height (Basic)}} & \multicolumn{2}{c}{\textbf{Sim-Sorting by Height (Advanced)}} & \multicolumn{2}{c}{\textbf{Sim-Sorting by Weight}} \\
\cmidrule{2-7}
 & Pre \,/\, Rec \,/\, F1 (\%) & eTaP \,/\,eTaR \,/\,eTaF1 (\%) & Pre \,/\, Rec \,/\, F1 (\%) & eTaP \,/\,eTaR \,/\,eTaF1 (\%) & Pre \,/\, Rec \,/\, F1 (\%) & eTaP \,/\,eTaR \,/\,eTaF1 (\%) \\
\midrule[0.5pt]
ICS-AD~\cite{fung2024attributions} & 72.7 \,/\, 66.2 \,/\, 69.3 & 63.9 \,/\, 83.1 \,/\, 72.3
 & \textbf{92.0} \,/\, 75.2 \,/\, 82.7
 & 75.4 \,/\, 87.6 \,/\, 81.1
 & 79.6 \,/\, 76.3 \,/\, 77.9
 & 54.8 \,/\, 88.1 \,/\, 67.6 \\
USAD~\cite{audibert2020usad} & 36.9 \,/\, 20.3 \,/\, 26.2 & 24.0 \,/\, 60.2 \,/\, 34.3
 & 41.6 \,/\, 34.9 \,/\, 38.0
 & 27.0 \,/\, 67.5 \,/\, 38.6
 & 35.3 \,/\, 31.7 \,/\, 33.4
 & 40.5 \,/\, 65.9 \,/\, 50.2 \\
OmniAnomaly~\cite{su2019robust} & 90.6 \,/\, 91.6 \,/\, 91.1 & 83.4 \,/\, 95.8 \,/\, 89.2
 & 58.5 \,/\, 92.8 \,/\, 71.8
 & 66.8 \,/\, 96.4 \,/\, 78.9
 & 80.3 \,/\, 92.9 \,/\, 86.1
 & 65.1 \,/\, 96.5 \,/\, 77.7 \\
MSCRED~\cite{zhang2019deep} & 35.6 \,/\, 78.9 \,/\, 49.1 & 24.9 \,/\, 89.5 \,/\, 39.0
 & 54.1 \,/\, 69.0 \,/\, 60.7
 & 32.2 \,/\, 84.5 \,/\, 46.6
 & 59.8 \,/\, 76.7 \,/\, 67.2
 & 32.2 \,/\, 88.3 \,/\, 47.2 \\
VAE-LSTM~\cite{lin2020anomaly} & 37.2 \,/\, 22.9 \,/\, 28.4 & 14.2 \,/\, 61.5 \,/\, 23.1
 & 23.3 \,/\, 16.0 \,/\, 19.0
 & 9.5 \,/\, 58.0 \,/\, 16.3
 & 41.6 \,/\, 65.4 \,/\, 50.9
 & 19.3 \,/\, 82.7 \,/\, 31.3 \\
GDN~\cite{deng2021graph} & 47.8 \,/\, 98.1 \,/\, 64.3 & 42.7 \,/\, 99.1 \,/\, 59.7
 & 70.4 \,/\, 89.7 \,/\, 78.9
 & 49.9 \,/\, 94.9 \,/\, 65.4
 & 59.5 \,/\, 99.2 \,/\, 74.4
 & 44.3 \,/\, 99.6 \,/\, 61.3 \\
MTAD-GAT~\cite{zhao2020multivariate} & 71.9 \,/\, 76.9 \,/\, 74.3 & 73.3 \,/\, 88.5 \,/\, 80.2
 & 67.4 \,/\, 50.8 \,/\, 57.9
 & 57.3 \,/\, 75.4 \,/\, 65.1
 & 77.2 \,/\, 86.0 \,/\, 81.4
 & 61.9 \,/\, 93.0 \,/\, 74.3 \\
GTA~\cite{chen2021learning} & 68.2 \,/\, 36.9 \,/\, 47.9 & 64.8 \,/\, 68.5 \,/\, 66.6
 & 91.2 \,/\, 58.4 \,/\, 71.2
 & 67.6 \,/\, 79.2 \,/\, 72.9
 & 87.7 \,/\, 50.2 \,/\, 63.9
 & 76.2 \,/\, 75.1 \,/\, 75.7 \\
LLMTIME~\cite{gruver2023large} & 35.9 \,/\, 61.8 \,/\, 45.4 & 41.2 \,/\, 80.9 \,/\, 54.6
 & 34.9 \,/\, 46.8 \,/\, 40.0
 & 36.2 \,/\, 73.4 \,/\, 48.5
 & 61.6 \,/\, 72.4 \,/\, 66.6
 & 25.7 \,/\, 86.2 \,/\, 39.6 \\
Sigllm~\cite{alnegheimish2024large} & 36.9 \,/\, 60.4 \,/\, 45.8 & 28.5 \,/\, 80.2 \,/\, 42.1
 & 52.5 \,/\, 73.4 \,/\, 61.2
 & 43.0 \,/\, 86.7 \,/\, 57.5
 & 67.4 \,/\, 84.9 \,/\, 75.2
 & 34.8 \,/\, 92.5 \,/\, 50.6 \\
JumpStarter~\cite{ma2021jump} & 64.7 \,/\, 80.5 \,/\, 71.7 & 72.0 \,/\, 90.3 \,/\, 80.1
 & 74.7 \,/\, 96.7 \,/\, 84.3
 & 61.3 \,/\, 98.4 \,/\, 75.5
 & 77.9 \,/\, 91.1 \,/\, 84.0
 & 70.4 \,/\, 95.6 \,/\, 81.1 \\
GeCos~\cite{wolsing2025gecos} & 81.2 \,/\, 98.9 \,/\, 89.2 & 78.3 \,/\, 99.5 \,/\, 87.6
 & 70.5 \,/\, \textbf{99.7} \,/\, 82.6
 & 32.5 \,/\, \textbf{99.8} \,/\, 49.0
 & 74.2 \,/\, \textbf{99.3} \,/\, 84.9
 & 58.3 \,/\, \textbf{99.7} \,/\, 73.6 \\
\midrule[0.5pt]
 IstGPT & \textbf{94.6} \,/\, \textbf{99.1} \,/\, \textbf{96.8} & \textbf{92.8} \,/\, \textbf{99.6} \,/\, \textbf{96.1}
 & 81.3 \,/\, 98.6 \,/\, \textbf{89.1}
 & \textbf{75.6} \,/\, 99.3 \,/\, \textbf{85.9}
 & \textbf{96.2} \,/\, 98.6 \,/\, \textbf{97.4}
 & \textbf{82.5} \,/\, 99.3 \,/\, \textbf{90.1} \\
\bottomrule[1pt]
\end{tabular}
\label{tab:Sim_datasets}
}
\end{table*}

%% file: table/Experiment/real-world.tex
\begin{table}[tbp]
\centering
\caption{Performance comparison on Real-world datasets.}
\resizebox{\linewidth}{!}{
\begin{tabular}{ccccccc}
\toprule[1pt]
{\multirow{2}{*}{\diagbox[width=8em]{\textbf{Models}}{\textbf{Datasets}}}} & \multicolumn{6}{c}{\textbf{Real-World}} \\
\cmidrule{2-7}
 & Pre (\%) & Rec (\%) & F1 (\%) & eTaP (\%) & eTaR (\%) & eTaF1 (\%)\\
\midrule[0.5pt]
ICS-AD & 67.8 & 85.5 & 75.6 & 64.3 & 82.6 & 72.3 \\
USAD & 85.3 & 67.4 & 75.3 & 67.8 & 75.2 & 71.3 \\
OmniAnomaly & 65.6 & 97.9 & 78.6 & 74.1 & 88.7 & 80.8 \\
MSCRED & \textbf{95.5} & 52.4 & 67.7 & \textbf{90.3} & 43.1 & 58.4 \\
VAE-LSTM & 89.5 & 65.8 & 75.8 & 63.9 & 71.4 & 67.5 \\
GDN & 61.2 & 93.1 & 73.9 & 57.6 & 79.2 & 66.7 \\
MTAD-GAT & 73.7 & 68.9 & 71.2 & 78.4 & 70.3 & 74.1   \\
GTA & 65.7 & 71.0 & 68.2 & 35.6 & 55.5 &  43.4 \\
LLMTIME & 58.9 & 60.6 & 59.7 & 27.8 & 54.9 & 36.9 \\
Sigllm & 48.7 & 91.2 & 63.5 & 21.9 & 70.5 & 33.4 \\
JumpStarter & 84.2 & 74.5 & 79.1 & 79.6 & 80.0 & 79.8 \\
GeCos & 93.8 & 40.8 & 56.9 & 79.5 & 13.8 & 23.6 \\
\midrule[0.5pt]
IstGPT & 89.5 & \textbf{96.7} & \textbf{93.0} & 83.7 & \textbf{93.2} &  \textbf{88.2} \\

\bottomrule[1pt]
\end{tabular}
\label{tab:real_world}
}
\end{table}

%% file: table/Experiment/time-cost.tex
\begin{table}[tbp]
\centering
\captionsetup{font={bf}}
\caption{The time cost of \sysname}
\label{tab:tab_time_cost}
\resizebox{\columnwidth}{!}{
\begin{tabular}{c|c|c|c|c|c|c}
\toprule[1pt]
{\diagbox[width=8em]{\textbf{Datasets}}{\textbf{Phases}}} & \textbf{DataProcess} & \begin{tabular}[c]{@{}c@{}}\textbf{ISTG-}\\\textbf{Gen}\end{tabular} & \begin{tabular}[c]{@{}c@{}}\textbf{ISTG-}\\\textbf{Learning}\end{tabular} & \begin{tabular}[c]{@{}c@{}}\textbf{Anomaly-}\\\textbf{Detection}\end{tabular} & \textbf{Total} & \begin{tabular}[c]{@{}c@{}}\textbf{Online}\\\textbf{Throughput}\end{tabular}\\
\midrule[0.5pt]
SWaT & 34s & 3.8m & 34m & 18s & 38.7m & 7195 sps \\
WADI & 73s & 6.7m & 75.8m & 27s & 84.2m &  6400 sps\\
Simulation & 2s & 50s & 8.3m & 8s & 9.3m & 450 sps\\
Real-world & 1s & 30s & 1.7m & 5s & 2.3m & 200 sps\\
\bottomrule[1pt]
\end{tabular}
}
\label{tab:detectDr}
\end{table}

%% file: table/Experiment/cost_comp.tex
\begin{table}[tbp]
\centering
\captionsetup{font={bf}}
\caption{Comparison of the average time costs for different baselines on various datasets}
\label{tab:tab_time_cost_cmp}
\resizebox{0.5\textwidth}{!}{
\begin{tabular}{c|c|c|c|c}
\toprule[1pt]
{\diagbox[width=8em]{\textbf{Models}}{\textbf{Datasets}}} & \textbf{SWaT} & \textbf{WADI} & \textbf{Simulation} & \textbf{Real-world} \\
\midrule[0.5pt]
ICS-AD & 1.7h & 4.6h & 3.4m  & 2.5m\\
USAD & 7.2h & 10.1h & 4.3m & 1.4m\\
OmniAnomaly & 7.4h & 11.1h & 5.6m & 2.1m\\
MSCRED & 14.2h & 18.3h & 16.8m & 10.1m \\
VAE-LSTM & 9.9h & 12.9h & 13.4m & 4.6m \\
GDN & 1.4h & 1.9h & 1.1m & 31s \\
MTAD-GAT & 9.4h & 14.8h & 15.4m & 5.2m \\
GTA & 3.3h & 9.3h & 13.9m & 3.4m \\
LLMTIME & 10.3h & 21.3h & 6.1m & 1.4m\\
Sigllm & 23.3h & 1.6d & 1.0h & 18.3m\\
JumpStarter & 27.5m & 1.2h & 2.1m & 1.1m\\
GeCos & 14.1h & 3.2d & 2.4m & 11s\\
\midrule[0.5pt]
IstGPT & 38.7m & 1.4h & 9.3m & 6.3m \\
\bottomrule[1pt]
\end{tabular}
}
\end{table}

%% file: tex/Discussion.tex
\sysname demonstrates strong accuracy and scalability for industrial anomaly detection. We now discuss several directions for future research.

\noindent
\textbullet{}
Effective deployment of \sysname relies on industrial multi-modal knowledge, which corresponds to standard engineering artifacts commonly maintained in modern ICS environments (e.g., system documentation and control logic).
Such knowledge is typically collected during system design and commissioning, and thus represents a one-time, reusable asset that enables largely automated anomaly detection with minimal additional human intervention.

\noindent
\textbullet{} By detecting abnormal sensors or actuators, \sysname can support timely mitigation actions (e.g., component isolation), thereby substantially reducing potential damage. 
However, due to inherent CPS constraints such as delayed observability and limited in-process intervention, some attack effects may still occur before detection~\cite{asghar2019cybersecurity, hofer2021industrial}.
Future work will explore tighter integration with complementary defense mechanisms, including resilient network architectures~\cite{qin2024cgan} and predictive maintenance strategies~\cite{nankya2023securing}, to further reduce end-to-end attack impact in ICS environments.

\noindent
\textbullet{} 
\sysname primarily detects realistic industrial attack scenarios that introduce observable spatiotemporal inconsistencies. 
An advanced, sophisticated adversary with knowledge of the dependency graph could potentially perform attacks that resemble normal causal behavior. 
However, executing such graph-aware false data injection attacks in practice is challenging because the attacker must simultaneously satisfy tightly coupled constraints on node validity, edge consistency, and logical coherence. 
As system scale increases, maintaining stealthy manipulations within the normal operating manifold becomes increasingly difficult, thereby significantly increasing the attack cost and making such attacks much harder in practice.

%% file: tex/RelatedWork.tex
\textbf{Industrial Sensors and Actuators Anomaly Detection.}
Existing approaches for industrial anomaly detection can be broadly categorized into three classes.

(1) \emph{Traditional rule-based methods.}
JumpStarter~\cite{ma2021jump} applies compressed sensing to detect anomalies by comparing reconstructed signals with original observations.
RRCF~\cite{guha2016robust} and SPOT~\cite{siffer2017anomaly} rely on extreme value theory to identify rare events by modeling statistical tails in data streams.
However, most rule-based methods struggle with high-dimensional and multivariate industrial data.
Although approaches such as GeCo~\cite{wolsing2025gecos} attempt to address this challenge through complex rule modeling, they are often time-consuming, unstable, and difficult to scale, limiting their robustness in realistic industrial environments.

(2) \emph{Sequence representation learning-based methods.}
USAD~\cite{audibert2020usad} employs adversarial autoencoders to detect anomalies using weighted reconstruction errors.
OmniAnomaly~\cite{su2019robust} leverages stochastic recurrent neural networks to learn latent representations and flags anomalies when reconstruction probabilities drop below a threshold.
Despite their effectiveness in temporal modeling, these methods lack explicit representations of inter-variable dependencies, which restricts their performance in structured industrial systems with strong sensor–actuator coupling.

(3) \emph{Graph learning-based methods.}
GDN~\cite{deng2021graph} detects anomalies by forecasting each sensor value based on its neighbors in a learned dependency graph.
MTAD-GAT~\cite{zhao2020multivariate} captures both feature-wise and temporal dependencies using graph attention mechanisms.
PGRGAT~\cite{wu2023physics} further incorporates physical constraints into graph attention networks to improve anomaly detection.
While these approaches explicitly model inter-variable relationships, they often incur high computational overhead due to data-driven dependency learning, limiting their practicality in large-scale industrial deployments.

\textbf{LLM-based Anomaly Detection.}
Recent studies explore large language models (LLMs) for time-series anomaly detection.
LLMTIME~\cite{gruver2023large} formulates forecasting as next-token prediction over text-encoded numerical sequences, enabling zero-shot detection.
SigLLM~\cite{alnegheimish2024large} identifies anomalies by measuring discrepancies between predicted and observed signals.
LLMAD~\cite{liu2024large} leverages in-context learning and chain-of-thought prompting to retrieve similar patterns for anomaly detection and interpretation.
However, these methods primarily focus on univariate signals and overlook inter-variable dependencies, resulting in limited accuracy for complex industrial processes.
In contrast, our work pioneers the use of LLMs for \emph{dependency graph construction}, enabling accurate and interpretable modeling of industrial sensor-actuator relationships.

%% file: tex/Conclusion.tex
We present \sysname, the first LLM- and graph learning-based industrial anomaly detection tool that constructs interpretable spatial-temporal graphs to detect anomalies in real time. \sysname leverages a multi-stage prompt engineering pipeline to extract sensor-actuator dependencies from industrial multi-modal knowledge, forming accurate spatial graphs. It then models spatial-temporal patterns via graph neural networks and an improved encoder-decoder architecture to detect anomalies through reconstruction errors. To ensure graph accuracy, \sysname introduces LLM-Optimation to iteratively validate and refine graph structure from the perspective of node accuracy, edge consistency, and logical coherence. Extensive experiments on 9 datasets, including public, simulation-based, and real-world cases, demonstrate that \sysname achieves high detection performance and outperforms 12 state-of-the-art baselines. These results highlight \sysname's effectiveness, scalability, and feasibility in practical industrial scenarios.